%% file: Karachentsev.tex
\documentclass[referee]{mn2e}
\usepackage{graphicx}%

\title[Emission Regions in the Local Volume Dwarf Galaxies]
{\bf{}Statistics and Properties of Emission-Line Regions in the Local Volume Dwarf Galaxies}

\author[I.\,D.\,Karachentsev, S.\,S.\,Kaisin]
{I.\,D.\,Karachentsev$^{1}$\thanks{E-mail:ikar@sao.ru},
{S.\,S.\,Kaisin}$^{1}$\thanks{E-mail:skai@sao.ru}\\
$^{1}$Special Astrophysical Observatory of the Russian Academy
 of Sciences, Nizhnij Arkhyz, KChR, 369167, Russia}
\begin{document}



\maketitle


\begin{abstract}
 We used the $H\alpha$ images from a large sample of nearby late-type dwarf galaxies to investigate properties of their emission structure. The sample consists of three hundred galaxies of the irregular (Irr), Magellanic irregular (Im), blue compact dwarf (BCD), and transition (Tr) types situated within a distance of 11 Mpc. In each galaxy, we indicated: the number of compact $HII$-regions, the presence of bubble-like or filament-like structures, the presence of a faint diffuse emission, and a sign of the global burst. The larger luminosity of a galaxy, the greater number of compact $HII$-sources in it. The integral and specific star-formation rates of the dwarf increase steeply with the increase of the number of $HII$-regions showing the evidence of the epidemic character of star-formation process. The dwarf galaxies with emission-line bubbles, or filaments, or signs of the global star-formation burst have approximately the same hydrogen-mass-to-luminosity ratio as that of the whole sample objects. However, their mean star-formation rate is significantly higher than that of other galaxies in the sample. Emission bubble-like structures are found in the nearby dwarfs with a frequency of 1 case per 4-5 galaxies. Their linear diameters are close to those expected for supernova remnants. The mean specific $SFR$ for the nearby late-type dwarfs is close to the Hubble parameter, $H_0 = -10.14$ dex (yr)$^{-1}$, consistent with the sluggish cosmic star-formation history of galaxies of this kind.

\end{abstract}


\section{Introduction}
Many emission-line regions observed in spiral galaxies usually are located orderly outlining the spiral structure of galaxies. Parameters of spiral pattern in galaxy discs have been the focus of research for many authors. In dwarf galaxies of stellar masses of $M^*<10^9 M_{\odot}$, the location of emission-line regions seems chaotic, since it is governed rather by turbulent gas motions than by the regular rotation of a galaxy. For this reason, systematic studies of the emission-line structures in a population of low-mass galaxies are rather rare. Authors usually are confined to measuring the integral $SFR$ of a galaxy from its observed flux in the $H\alpha$ emission line, or by analysing the structure and kinematics of $HII$ regions in selected dwarf objects.

At present, the most representative sample of dwarf galaxies is the Updated Nearby Galaxy Catalog (=UNGC, Karachentsev et al. 2013) which comprises more than 800 galaxies with distances within 11 Mpc. Other notable surveys of nearby dwarfs include Rosenberg et al. 2006, Kennicutt et al. 2008, Dale et al. 2009, and Salzer et al. 2000. Dwarf galaxies are about 5/6 of the population of this sample. The refined and updated version of the UNGC catalogue is available in the data base: http://www.sao.ru/lv/lvgdb. Over the past decade, systematic observations of the Local Volume galaxies (=LV, $D<11$ Mpc) in the  $H\alpha$ emission line and in the red continuum have been carried out with the 6-m BTA telescope of the Special Astrophysical Observatory of the Russian Academy of Sciences (SAO RAS). These observations are described in a number of papers including the most recent (Karachentsev et al. 2015, Kaisin \& Karachentsev 2019) and references therein. The obtained $H\alpha$ images of nearby galaxies are also available in the aforementioned data base. A significant part of these galaxies have accurate distances measured from the luminosity of the tip of the red giant branch at the Hubble Space Telescope.  

We considered the $H\alpha$ images of dwarf galaxies from our data base for the purpose of classification of the structure of emission-line regions in them. Our collection was increased with the $H\alpha$ images of LV dwarfs that we found in publications of other authors:  Hunter et al. 1993, Hodge \& Miller 1995, Shulte-Ladbeck \& Hopp 1998, van Zee 2000, Skillman et al. 2003, James et al. 2004, Bouchard et al. 2009, Cote et al. 2009, Lee et al. 2009, and McQuinn et al. 2019. For analysis, we selected only late-type dwarf systems: irregular -Ir, Magellanic irregular - Im, blue compact dwarf - BCD, and transition - Tr disregarding the subsystem of spheroidal (dSph) and elliptical (dE) dwarf galaxies which as a rule do not show any $H\alpha$ emission. Here we should clarify: there is the known case of the classical dSph galaxy DDO~44, where a faint emission-line region illuminated by several blue stars has been detected (Karachentsev et al. 2011).

The considered sample contains the $H\alpha$ images of 309 LV dwarfs. The data on them are given in Appendix.

\section{Basic Parameters of the Local Volume Dwarfs}
The summary of late-type dwarf galaxies (Appendix) includes the following characteristics: (1) --- name of a galaxy; (2) ---  morphological type; (3) ---  distance in Mpc; (4) ---  linear diameter (in kpc) measured within a Holmberg isophote of $26.5^m/\sq\arcsec$  in the $B$-band; (5) --- the average surface brightness in the $B$-band within the Holmberg diameter; (6) ---  logarithm of the integral luminosity of a galaxy in the $K$-band in solar-luminosity units; as known, $K$- band photometry is usually poorly constrained for dwarf galaxies with the 2MASS survey (Jarrett et al. 2000; 2003); actually, only a quarter of galaxies in our sample have $K_s$-data from 2MASS; for the remaining 3/4 dwarf galaxies their $K$-magnitudes were determined via apparent $B$-magnitudes and the average color index $<B-K> = 2.35$ according to Jarrett et al. 2003; under such approach a typical error in $L_K$- luminosity is about 0.2 dex;(7)~---  logarithm of the neutral hydrogen mass in solar-mass units; (8) ---  logarithm of the integral star-formation rate ($M_{\odot}$/yr) determined according to Kennicutt (1998) as
                 
                 $$\log[SFR]_{H\alpha}  = \log F_c(H\alpha)+2\log D +8.98,$$
where $D$ is the distance (in Mpc) and $F_c(H\alpha)$ is the integral flux in the $H\alpha$ line corrected for the internal and Galactic extinction; (9) ---  logarithm of the integral star-formation rate ($M_{\odot}$/yr) determined from the ultra-violet flux in the $FUV$-band, $F_c(FUV)$, corrected for the extinction according to Lee et al. (2011),
                
                $$\log[SFR]_u  = \log F_c(FUV)+2\log D -6.78.$$ 
Here, Galactic extinction in the $B$-band is accounted according to Schlegel et al. 1998. For internal extinction in the $B$-band we use the expression by Verheijen, 2001

                $$A_{i,B} = [1.54 +2.54(\log V_m - 2.2)] \log (a/b) $$,
if $V_m > 39$ km/s, otherwise $A_{i,B}$ = 0, where $V_m$ is a rotation amplitude of the galaxy and $(a/b)$ is its apparent axial ratio. For galaxies unobserved in $HI$ we adopt $A_{i,B}$ = 0; their fraction in our sample is 16\%. To correct the extinction in the $H\alpha$-line and in the $FUV$-band 
we use the relations:

                $$ A_{H\alpha} = 0.54 A_B  \,\, {\rm and} \,\, A_{FUV} = 1.93 A_B. $$ 
Note, that the adopted internal extinction in our sample dwarf galaxies is very small, being at average about 0.01 mag.

All the above parameters given in the Appendix are taken from the latest and extended version of the UNGC catalogue (http://www.sao.ru/lv/lvgdb).

The distribution of late-type dwarf galaxies by the $K$-luminosity and linear diameter, as well as by the hydrogen mass and the diameter is presented in the left-hand and right-hand panels of Fig. 1, respectively. The galaxies of  Irr and Im morphological types are indicated by the solid circles, blue compact dwarfs (BCDs) are shown by the asterisks, and transition-type systems (Tr) are indicated by the empty circles. As one can see, the population of LV dwarfs is characterized by diameters in the range of [0.1--10] kpc, $L_K$-luminosities in the interval of $[10^5 - 2\times10^9] L_{\odot}$, and hydrogen masses of $[10^5 -10^9]M_{\odot}$. Note that for several ultra-faint dwarfs, only the upper limit of the $HI$ flux was used as an estimate of the hydrogen mass. Judging from the slope of the regression lines (the last column in Table 1), the bulk of dwarf systems follow the lines of the constant surface density: $L_K/A^2_{26}$ and $M_{HI}/A^2_{26}$.  Transition-type objects are systematically located below the regression lines. BCD dwarfs are characterized by a higher surface brightness than Irr and Im galaxies.

The left-hand and right-hand panels of Fig. 2 present the distribution of the integral and specific (per unit of the $L_K$ luminosity) star-formation rates determined from the $H\alpha$ flux depending on luminosity. Galaxies of different types are indicated with the same symbols as in Fig.1. In some cases, to estimate the $SFR$, we used the upper limit of  $H\alpha$ flux of a galaxy. Transition-type dwarf galaxies stand out from the rest by low values of both the integral and specific star-formation rate. In general, the population of LV dwarfs approximately follows the line of constant specific star-formation rate, although, individual objects vary in this parameter by several orders. Note that the regression line in the right-hand panel of Fig. 2 is drawn over the entire set of LV dwarfs. Elimination of transition-type dwarfs with unreliable $SFR(H\alpha$) estimates reduces the slope of the regression line almost to zero.
The mean value for LV dwarfs, $\langle SFR/L_K\rangle = -10.55\pm0.04$, means that for a cosmological time of 13.8 Gyr, a galaxy with the ratio of stellar mass-to-$K$-luminosity $M^*/L_K= 1.0M_{\odot}/L_{\odot}$ (Bell et al. 2003) manages to reproduce about 39\% of its stellar mass. According to McGaugh \& Schombert (2014) and Zhang et al. (2017), in the case $M^*/L_K = (0.4-0.6)M_{\odot}/L_{\odot}$ and with the steady star-formation rate, a typical LV dwarf can reproduce 65 -- 100\% of its stellar mass. In other words, the characteristic star-formation rate of late-type dwarf galaxies was only marginally more intensive in the past than the current rate.

The left-hand panel of Fig. 3 shows the distribution of objects from our sample by the integral star-formation rate and mean surface brightness. The LV dwarf galaxies vary by five magnitudes in surface brightness. Fainter galaxies show lower $SFRs$. This effect is partially due to the known correlation between luminosity and surface brightness of a galaxy. However, as is seen from the right-hand panel of Fig.~3, the specific star-formation rate, $SFR/L_K$, also shows a tendency to decrease from high surface-brightness objects to faint galaxies. One of the most noticeable ``deviating'' objects in these diagrams is the ``Garland'' tidal dwarf system which looks like a chain of emission knots on the outskirts of NGC~3077. \footnote{For faint dwarfs with $SB_B >26.5^m/\sq\arcsec$, the average surface brightness was estimated within the effective radius but not within the Holmberg isophote.}

The $SFR$ estimation via the $H\alpha$-line flux refers to a characteristic time scale of $\sim$10 Myr. The flux in the FUV band determines the star-formation rate on a time scale of $\sim$100 Myr. Comparison of $SFR(H\alpha$) and $SFR(FUV)$ allows us to detect galaxies that are in different stages of star-forming activity. The observed data on $SFR(H\alpha$) and $SFR(FUV)$ for the LV galaxies have been discussed in publications by Lee et al. (2009), Lee et al. (2011) and Karachentsev \& Kaisina (2013). Figure~4 presents the ratio between $SFR(H\alpha$) and $SFR(FUV)$ for the dwarf sample under consideration. Notations for the galaxies of different types are shown with the same symbols as in the previous figures. The closest correlation occurs for BCD galaxies. Transient dwarf systems are located in the region of the minimum values, where the $SFR$-determination error is rather large. The regression line in the figure specifies the well-known fact that in low-mass galaxies the $SFR(H\alpha$) estimates are systematically slightly lower in comparison with the $SFR(FUV)$ estimates. Figure~5 shows the ratio of these values for galaxies with different luminosities. A weak tendency of the decrease of the $SFR(H\alpha)/SFR(FUV)$ ratio with the decrease of the galaxy's luminosity can be seen in this figure. In general, the dispersion of this ratio grows towards faint galaxies, which may indicate a more significant role of star-forming bursts in the lowest-mass galaxies. The diagram contains the objects (DDO~120, KDG~52, NGC~4656UV), whose star-forming activity has decreased by more than an order of magnitude over the last $\sim$10 Myr. A few galaxies with the ratio $SFR(H\alpha)/SFR(FUV)>3$ are obviously in the stage of current star-formation burst.

Figure 6 reproduces the images of two dwarf galaxies, DDO~120 and Mrk~475, staying in significantly different star-formation stages. The left-hand images of galaxies in the $FUV$ band were taken from the GALEX survey (Martin et al. 2005), http://galex.stsci.edu/Galex/View; middle and right-hand images were obtained with the 6-meter BTA telescope in the $H\alpha$ line (continuum-subtracted) and in the red continuum. The image size is $4^{\prime}\times4^{\prime}$. In the case of DDO~120, the $SFR(H\alpha)/SFR(FUV)$ ratio is --2.20 dex, while for the galaxy Mrk~475, this ratio reaches a value of +0.74 dex.

Two scenarios are possible in the pattern of burst activity in dwarf galaxies: a) the burst results in the release of the galaxy's gas into the intergalactic space, b) the energy of  burst is small and the galaxy's gaseous component remains within its volume. Figure~7 presents the distribution of dwarf galaxies under study by the ratios $M_{HI}/L_K$ and $SFR(H\alpha)/SFR(FUV)$. The absence of significant correlation in this diagram may serve as an argument against the scenario of irrecoverable gas loss during a star-formation burst.

The linear regression parameters, y = A + Bx, for the diagrams presented above are given in Table~1. The table columns contain: N --  number of objects in the diagram, R -- correlation coefficient, SD -- standard deviation from the regression line, A and B -- the linear regression coefficients.

\section{Emission Properties and Morphology of the LV Dwarfs}
We divided the sample of 309 late-type dwarf galaxies with $H\alpha$ imaging into three categories: Irr+Im, BCD, and Tr. The relative number of objects in each category is 78\%, 16\%, and 6\%, respectively, which is approximately consistent with the overall occurrence of these types in the UNGC catalogue. Table~2 presents the mean values of various global parameters for dwarfs of different types including the standard error of the mean. The underlined values differ from the mean for the whole sample by more than 3$\cdot$SD. In these data, the following patterns can be distinguished.
\begin{itemize}
  \item Blue Compact Dwarfs (BCDs) have approximately the same linear diameters as the Irr+Im types but have twice as high $K$- luminosity. Their surface brightness is higher on average by 0$\fm$9. BCDs and Irr+Im dwarfs are characterized by approximately similar hydrogen-mass abundance, but the specific ratio $M_{HI}/L_K $ for BCDs appears to be three times smaller than that for Irr+Im systems. The mean integral star-formation rate for BCD galaxies is two-three times higher than that for Irr+Im galaxies; however, their average specific star-formation rates are approximately the same.

   \item Dwarf galaxies of the transition- type (Tr) are on average two times smaller than Irr+Im galaxies. The luminosity of Tr dwarfs is on average 1/8 of the luminosity of Irr+Im dwarfs, and their surface brightness is two times fainter than that of Irr+Im. The integral abundance of the hydrogen mass of Tr dwarfs is on average 30 times smaller than that of Irr+Im, and the $M_{HI}/L_K$ specific ratio is on average three times smaller, too. The integral star-formation rate for Tr galaxies is by two orders lower than that for Irr+Im, being near the sensitivity limit for the H$\alpha$ and FUV surveys. Their $SFR/L_K$ is at least an order lower than that for other dwarfs.
\end{itemize}

 Each galaxy in the UNGC catalog is supplied with a dimensionless ``tidal index'' $\Theta_1$ indicating a density contrast contributed by the most significant neighbour of the galaxy. The cases with negative $\Theta_1$ can be considered as well isolated (field) objects. Statistics of $\Theta_1$ presented in UNGC shows that the relative numbers of isolated objects (i.e. those with  $\Theta_1 < 0$) among dwarfs of various types are: $q$(Irr+Im) = $0.51\pm0.09$, $q$(BCD) = $0.78\pm0.12$, and $q$(Tr) = $0.06\pm0.06$. Thus, almost all transition-type dwarfs with accurately measured distances have 
$\Theta_1 > 0$, locating in sites around massive galaxies. Consequently, their small sizes, low stellar masses, hydrogen masses, and low star-formation rates can be the result of ``stripping'' of dwarf galaxies under their  motion in the halo of massive neighbours.

  The BCD-type dwarfs are, however, predominantly isolated objects. Their number in the Local Volume is five times smaller than the number of Ir+Im galaxies. If we assume that a part of the Irr+Im dwarfs transform into BCD objects during the term of star-formation burst, then the characteristic burst amplitude would be a factor of (2--2.5). We note that the limits of the existing galaxy surveys by fluxes in the $HI, H\alpha$ lines, and in the $FUV$ band are quite sufficient for detecting dwarf galaxies of the Irr, Im, and BCD types up to the the Local Volume edge. However, in order to reliably detect transition-type dwarfs, deeper surveys are necessary.

\section{Emission Knot Numbers in the LV Dwarfs}
Late-type dwarf galaxies differ from each other by a number of compact $HII$-regions. Data on the number of emission knots seen in the body of the LV dwarfs are presented in the 10-th column of the table in Appendix. On average, there are 2.6 compact emission knots per dwarf system. There is no evidence of compact knots in 22\% of the dwarfs in our sample. Table 3 contains the mean values of various global parameters of dwarf galaxies depending on the number, $n$, of emission knots in them: 0, 1, 2+3, $>3$. The following conclusions can be drawn from the analysis of these data. 
\begin{itemize}
\item The increase in the number of knots, $n$, is accompanied by the increase of the mean values of the linear diameter, the stellar mass ($K$-luminosity), and the hydrogen mass of the parent galaxy.

 \item In dwarf galaxies without emission knots, the average surface brightness is smaller by $0\fm5-0\fm6$  relative to the galaxies with knots.

 \item The ratio of hydrogen mass to stellar mass grows slightly with the increase in the number of knots, which is not consistent with the concept of gas depletion while forming new sources of star formation.

 \item For dwarf galaxies without emission knots, the mean star-formation rate is --4.3 dex which is close to the detection threshold of the LV-galaxy flux in $H\alpha$ and $FUV$.

 \item The characteristic star-formation rate for dwarfs with a single emission knot is --3.2 dex. Approximately the same integral star-formation rate is observed for single intergalactic emission knots (``$H\alpha$ sparkles'') in the space of the nearest group around M~81 (Karachentsev et al. 2011).
  
 \item The specific star-formation rate, $SFR/L_K$, systematically increases with the increase in the number of emission sites, which probably indicates the epidemic character of star formation in dwarf galaxies.

 \item The $SFR(H\alpha)/SFR(FUV)$ ratio shows a positive correlation with the number of star-formation sources. This effect is probably due to the known increase of the $SFR(H\alpha)/SFR(FUV)$ ratio with the increase of the dwarf galaxy luminosity.
\end{itemize}

\section{Texture of the Emission Regions}
Besides compact $HII$-regions, $H\alpha$ images of late-type dwarf galaxies show the presence of other emission structures. We divided them into four categories.

{\em B-bubbles.} Ring-like structures similar to supernova remnants (SNR).

{\em F-filaments.} One-dimensional structures of various extensions, which could be formed under the influence of shock waves in the interstellar medium.

{\em D-diffuse emission.} Faint amorphous luminescence in the presence or absence of compact $HII$-regions. The diffuse emission can be supported by B5--B9 stars that aged after a star-formation burst.

{\em G-global emission.} Bright $H\alpha$ emission covering the main galaxy body. This feature obviously indicates the fact that a dwarf galaxy is in the active stage of a star-formation burst.

The last four columns of the table in Appendix give the data on the presence of these features in our sample galaxies. Figure~8 shows the examples of galaxies with the specified properties: UGC~7427 (B), DDO~125 (F), dw1343+58 (D), and UGC~6456 (G). The right- and left-hand images of each galaxy correspond to the images in the red continuum and in the $H\alpha$ line, respectively. The image size is $4\arcmin\times4\arcmin$. 

The mean global parameters of dwarf galaxies with various details of the emission structure are given in Table~4. The mean values for each subset that differ from the average for the whole sample by more than 3$\cdot$SD are underlined. The relative occurrence of each emission detail is: 17\% (B), 4\% (F), 33\% (D), and 19\% (G). Comparison of the mean values of various parameters allows us to note the following specific characteristics.
\begin{itemize}

\item Dwarf galaxies with ring structures do not stand out among others in the linear sizes, integral luminosity, and hydrogen-mass abundance. However, their integral and specific star-formation rates are 2--3 times higher than the average.

\item Galaxies with a filament structure have larger linear sizes and higher $K$-luminosity. The integral star-formation rate of these dwarfs is by an order magnitude higher than the average for the whole sample.  Their observed specific star-formation rate, --9.96 dex (yr)$^{-1}$, exceeds the value of the Hubble parameter, $H_0 = -10.14$ dex (yr)$^{-1}$.
\end{itemize}

The most illustrative examples of the galaxies with filament structures are NGC~1569 and NGC~4605, whose $H\alpha$ images are shown in Fig.~9. Their distances and $L_K$-luminosities are 3.19 Mpc, 9.40 dex (NGC~1569) and 5.55 Mpc, 9.70 dex (NGC~4605), respectively. However, both the galaxies are classified as Sm-type spirals and, therefore, are not included in the dwarf sample under consideration.
\begin{itemize}
\item Dwarf galaxies with a diffuse emission do not distinguish themselves among other dwarfs in most parameters. At the same time, their integral star-formation rate is two times higher than the average for the whole sample.

\item Dwarf systems with a global star-formation burst have a higher surface brightness, as well as the integral star-formation rate, which is 3--10 times higher than the average for the whole sample. These galaxies are characterized by the highest specific star-formation rate, --9.83 dex, and the average ratio $SFR(H\alpha)/SFR(FUV)>1$. The last property indicates a significant increase in the star-formation process in these galaxies over the last $\sim10$ Myr. As was noticed above, only 1/5 of the late-type dwarf galaxies are at the burst phase, at which the star-formation rate several times increases.
\end{itemize}
The statistics on the environment influence on bursts in dwarf galaxies shows that this effect is of secondary importance (Karachentsev \& Kaisina 2013).  According to Skillman (2005), McQuinn et al. (2009), Stetson et al. (2007), and Emami et al. (2019), the gas-to-star conversion in low-mass galaxies has the character of stochastic oscillations due to particularly internal reasons.

\section{Bubbles as Probable Supernova Remnants}
Among the 309 LV dwarf galaxies considered, there are 52 objects in which one or more emission bubble-like structures are observed. Their typical angular sizes are $5-10\arcsec$, which, with an average distance of $\sim5$ Mpc, corresponds to a linear diameter of $\sim120-240$ pc. Such sizes of the observed bubbles are comparable to the characteristic diameter of supernova remnants, $\sim100$ pc.

Courtes et al. (1987) performed an $H\alpha$ survey of the neighbouring spiral galaxy M~33 and found 50 bubble-like regions in it. The integral luminosity of M~33 in the $K$-band is $L_K=4.1\times10^9 L_{\odot}$. To quantify the frequency of visible SNRs in a galaxy, one can introduce a parameter 
$\lambda_K$ meaning a specific luminosity of the galaxy per one supernova remnant. In the case of M~33 it is $\lambda_K=0.8\times10^8 L_{\odot}$.
The total luminosity of 309 dwarf galaxies equals $3.55\times 10^{10} L_{\odot}$. Assuming that the appearance of SNRs in a galaxy is a random event that depends only on the luminosity of a galaxy and scaling from the M~33 bubble statistics, we expect the total number of bubbles to be about 440 in the $H\alpha$ images of these galaxies. The actual number of bubbles in this ensemble of dwarf galaxies turned out to be 72, i.e., 6 times smaller yielding $\lambda_K=5.0\times10^8 L_{\odot}$. Deficiency of the observed SNR candidates is obviously caused by the more distant location of our galaxies, with which a significant part of SNRs are close to the angular resolution limit of the images we obtained.

In the recent literature, we have found estimates of the number of SNRs for several more distant galaxies: NGC~6946  $(L_K=9.8\times10^{10} L_{\odot}, D = 7.73$ Mpc, $N_{SNR}$ = 130, Long et al. 2019), M~83 ($L_K=7.2\times10^{10} L_{\odot}, D = 4.90$ Mpc, $N_{SNR}$ = 199, Williams et al. 2019), NGC~1569 ($L_K=2.5\times10^9 L_{\odot}, D = 3.19$ Mpc, $N_{SNR}$ = 10, Ercan et al. 2018), and NGC~2366 ($L_K=5.0\times10^8     L_{\odot}, D = 3.28$ Mpc, $N_{SNR}$ = 2, Vucetic et al. 2019). The mean specific luminosity of galaxy per one supernova remnant for them is five times higher than that for M~33, being    $\lambda_K=(4.0\pm1.2)\times10^8 L_{\odot}$. This quantity practically coincides with the estimate $\lambda_K=5.0\times10^8 L_{\odot}$, derived for our dwarf galaxy sample.

The distribution of dwarf galaxies by the number n of the observed bubbles in them approximately corresponds to the random Poisson distribution with the average value $\langle n\rangle=0.233$. However, there are four galaxies: Cas~1, Holmberg~II, DDO~131, and DDO~133 with n = 3 and 4, the random probability of which does not exceed $p\sim10^{-4}$ by the criterion $\chi^2$. The difference between the observed and the Poisson distribution is obviously due to the diversity of the considered dwarf galaxies in luminosities and distances.

It is curious to note that Evans et al. (2019) have discovered two weak emission ring-like structures with a diameter of $\sim80$ pc in the low-luminosity dwarf galaxy Leo~P $(L_K=7.7\times10^5 L_{\odot}, D=1.73$ Mpc).  If both of them were caused by independent supernova bursts, then the probability of this event would be extremely small. 

It should be emphasized that those bubble-like structures that we found are SNR candidates only. To clarify their physical nature, measurements of the line intensity ratios $H\alpha$/[NII] or $H\beta$/[OIII] are necessary. The project of such a program was described by Moumen et al. (2019).

\section{Concluding Remarks}
The Local Volume, in which dwarf galaxies predominate, favorably differs from other samples of low-mass galaxies in its completeness down to rather low luminosities of galaxies, as well as in the presence of accurate distance estimates for most dwarf systems. Our collection contains images in the $H\alpha$ emission line for 309 late-type dwarf galaxies with distances within 11 Mpc. We estimated average characteristics of dwarf galaxies: the linear diameter, the integral luminosity, the hydrogen mass, the integral and specific star-formation rates depending on the galaxy morphological type (BCD, Irr+Im, or Tr), the number of compact $HII$-regions, and the presence of various emission details of the structure (bubbles, filaments, diffuse emission, and a global burst).

The majority of our sample dwarfs (78\%) are of the Irr+Im type, half of which (51\%) belong to the field population, and the other half are resided in nearby groups. Blue compact dwarfs (BCDs) are 16\% of the sample under consideration and 78\% of them are isolated objects. Their integral luminosity and star-formation rate are significantly higher than those of the Irr, Im galaxies. The transition-type dwarf systems (Tr) make up only 6\% of the entire sample, and almost all of them (94\%) are located in the region of gravitational influence of neighbouring massive galaxies. Their small linear sizes and integral luminosities, low gas abundances, and low star-formation rates indicate that the structure of Tr dwarfs was significantly suppressed by a massive neighbour.

 About 78\% of the late-type dwarfs show the presence of one or more compact $HII$-regions. The larger the linear size and luminosity of a galaxy, the greater number of compact $HII$ sources in it. The integral and specific star-formation rates steeply increase with the increase of the number of $HII$-regions showing the evidence of the epidemic character of star formation. The average hydrogen-mass-to-luminosity ratio does not decrease with the increasing number of emission sites, which diverges from the concept of quick gas depletion during star formation. This feature can be caused by resupply of the galaxy gas from the intergalactic medium.

Dwarf galaxies, in the emission-line structure of which bubbles, filaments, or signs of the global star-formation burst are observed, have approximately the same ratio of hydrogen mass to luminosity as that of the whole sample considered. However, their mean star-formation rate is significantly higher than that of other galaxies in the sample.

Emission bubble-like structures are found in the LV galaxies with a frequency of 1 case per 4--5 galaxies. Their linear sizes are close to those expected for supernova remnants. Although, a significant part of the supposed SNRs in our sample is probably lost due to the low angular resolution of the $H\alpha$ images ($\sim2\arcsec$).

With the ratio $M^*/L_K=0.5 M_{\odot}/L_{\odot}$ (McGaugh \& Schombert 2014, Zang et al. 2017), the mean specific star-formation rate in late-type dwarf galaxies is --10.25 dex (yr)$^{-1}$ derived via the $H\alpha$ flux and --9.84 dex (yr)$^{-1}$ via the $FUV$-flux. Both estimates are close to the value of the Hubble parameter, $H_0= -10.14$ dex (yr)$^{-1}$. This means that the star-formation rate currently observed is quite sufficient for reproducing the observed mass of the galaxy over the cosmological time of 13.8 Gyr. In other words, star formation in late-type dwarf galaxies is obviously sluggish without any significant global acceleration or deceleration on the cosmological scale of $H_0^{-1}$. Short-term star-formation bursts caused by internal processes in a dwarf galaxy or by its merging with another dwarf system have little effect on the overall picture of the gas-to-star conversion.

Finally, it should be noted that the existing dwarf galaxy surveys in the $HI, H\alpha$ lines, and in the $FUV$ band are quite sufficient for detecting the majority of the LV late-type dwarfs. However, expanding the sample beyond 11 Mpc will require deeper sky surveys.

{\bf Acknowledgements}
 We are grateful to the anonymous referee for a report that helped us improve the manuscript.
This work is supported by RSF grant 19--12--00145.

\clearpage
\begin{table*}
\caption{Parameters of linear regression on Figures 1-5 and 7.}
\begin{tabular}{lcrcrr} \\ \hline  
 Relation                           &            N  &   R   &   SD    &     A    &           B\\
\hline
$\log(L_K)$ vs. $\log(A_{26})$     &            309 &  0.85 &  0.38  &  7.01$\pm$0.03 &  1.90$\pm$0.07\\

 $\log(M_{HI})$ vs. $\log(A_{26})$     &        261 &  0.76 &  0.47  &   6.86$\pm$0.04 &  1.78$\pm$0.09\\

 $\log(SFR_{H\alpha})$ vs. $\log(L_K)$    &     309 &  0.77 &  0.71 &  --11.78$\pm$0.42 &  1.16$\pm$0.06\\

 $\log(SFR)_{H\alpha}/L_K)$ vs. $\log(L_K)$  &  309 &  0.17 &  0.71 &  --11.77$\pm$0.42&   0.16$\pm$0.06\\

 $\log(SFR_{H\alpha})$ vs. $SB$          &      309 & --0.38 &  1.02 &   8.35$\pm$1.55 &  --0.46$\pm$0.06\\

 $\log(SFR_{H\alpha}/L_K)$ vs. $SB$       &     309 & --0.19 &  0.71 &  --7.01$\pm$1.07&  --0.14$\pm$0.04\\

 $\log(SFR_{H\alpha})$ vs. $\log(SFR_{FUV})$  & 254&   0.87&   0.55 &  --0.15$\pm$0.11 &  1.09$\pm$0.04\\

 $\log(SFR_{H\alpha}/SFR_{FUV})$ vs. $\log(L_K)$ &254 &  0.16 &  0.54 &  --1.33$\pm$0.37 &  0.12$\pm$0.05\\

 $\log(M_{HI}/L_K)$ vs. $\log(SFR_{H\alpha}/SFR_{FUV})$ &   214 &  0.04 &  0.53 &  -0.22$\pm$0.05 &  0.03$\pm$0.07\\
\hline
\end{tabular}
\end{table*}

\begin{table*}
\caption{Mean parameters of the LV dwarfs vs. their morphological type.}
\begin{tabular}{lrrrr} \\ \hline

Mean parameter     &   all      &   Irr,Im    &     BCD  &        Tr\\
\hline
Number            &    309     &     242    &       51   &        16\\
\hline
$\langle\log(A_{26}/kpc)\rangle$    &  0.29$\pm$0.02  & 0.31$\pm$0.02 &  0.28$\pm$0.05 &  \underline{0.01$\pm$0.13}\\
                                                          
$\langle\log(L_K/L_{\odot})\rangle$       & 7.56$\pm$0.04 &  7.55$\pm$0.05 &  \underline{7.92$\pm$0.09}&   \underline{6.63$\pm$0.13}\\
                                            
$\langle SB\rangle, ^m/\sq\arcsec$ &24.50$\pm$0.05  &24.60$\pm$0.06 & \underline{23.73$\pm$0.09} & \underline{25.35$\pm$0.13}\\
                                            
$\langle\log(M_{HI}/M_{\odot})\rangle$      & 7.43$\pm$0.04 &  7.52$\pm$0.04 &  7.37$\pm$0.11 &  \underline{5.96$\pm$0.20}\\
                                                          
$\langle\log(M_{HI}/L_K)\rangle$      &--0.22$\pm$0.03 & --0.11$\pm$0.03 & \underline{--0.63$\pm$0.07} & \underline{--0.66$\pm$0.15}\\
                                            
$\langle\log(SFR_{H\alpha})\rangle$      & --2.99$\pm$0.06 & --2.96$\pm$0.07 & \underline{--2.52$\pm$0.14} & \underline{--4.98$\pm$0.19}\\
                                           
$\langle\log(SFR_{FUV})\rangle$      &--2.69$\pm$0.05  &--2.64$\pm$0.06 & \underline{--2.30$\pm$0.10} & \underline{--4.38$\pm$0.15}\\
                                             
$\langle\log(SFR_{H\alpha}/L_K)\rangle$   &--10.55$\pm$0.04 & --10.51$\pm$0.04 &--10.44$\pm$0.08& \underline{--11.61$\pm$0.18} \\    
                                                         
$\langle\log(SFR_{H\alpha}/SFR_{FUV})\rangle$   &--0.41$\pm$0.03 & --0.43$\pm$0.04  &--0.26$\pm$0.07  &--0.60$\pm$0.20\\
\hline
\end{tabular}
\end{table*}

\begin{table*}
\caption{Mean parameters of the LV dwarfs vs. the number of compact HII-regions.}
\begin{tabular}{lrrrr} \\ \hline

 Mean parameter      &    n=0    &       n=1    &       n=2,3   &        n$>$3   \\     
\hline
 Number               &    69    &        68   &          86   &          86\\
\hline
$\langle\log(A_{26}/kpc)\rangle$            &  0.08$\pm$0.04 &    0.14$\pm$0.04  &   0.31$\pm$0.02  &   0.55$\pm$0.02\\

$\langle\log(L_K/L_{\odot})\rangle$         &  6.99$\pm$0.08&     7.32$\pm$0.09 &    7.64$\pm$0.05 &    8.13$\pm$0.06\\
 
$SB, ^m/\sq\arcsec $                          &   24.91$\pm$0.11 &   24.34$\pm$0.12 &   24.37$\pm$0.09 &   24.41$\pm$0.10\\

$\langle\log(M_{HI}/M_{\odot})\rangle$       &   6.77$\pm$0.09 &    7.11$\pm$0.09  &   7.49$\pm$0.06  &   7.97$\pm$0.05\\

$\langle\log(M_{HI}/L_K)\rangle$            &  --0.33$\pm$0.07  &  --0.27$\pm$0.07 &   --0.17$\pm$0.06 &   --0.16$\pm$0.06\\

$\langle\log(SFR_{H\alpha})\rangle$            &  --4.31$\pm$0.10  &  --3.23$\pm$0.11  &  --2.67$\pm$0.07 &   --2.07$\pm$0.06\\

$\langle\log(SFR_{FUV})\rangle$             &  --3.62$\pm$0.10 &   --2.86$\pm$0.09  &  --2.47$\pm$0.06  &  --1.93$\pm$0.05 \\ 

$\langle\log(SFR_{H\alpha}/L_K)\rangle$     & --11.30$\pm$0.08  & --10.55$\pm$0.09  & --10.31$\pm$0.05 &  --10.20$\pm$0.05 \\  

$\langle\log(SFR_{H\alpha}/SFR_{FUV})\rangle$           &  --0.74$\pm$0.09  &  --0.38$\pm$0.07  &  --0.33$\pm$0.05  &  --0.21$\pm$0.04\\
\hline
\end{tabular}
\end{table*}

\begin{table*} 
\caption{Mean parameters of the LV dwarfs for different emission-line structures.}
\begin{tabular}{lrrrrr} \\ \hline

 Mean parameter                       &    Bubble   &    Filament  &    Diffuse &      Global    &      All\\
\hline
 Number                               &       52    &        13   &        101     &     58    &        309\\
\hline 
$\langle\log(A_{26}/kpc)\rangle$      &   0.42$\pm$0.05 &   \underline{0.50$\pm$0.05} &   0.36$\pm$0.03 &   0.33$\pm$0.05 &   0.29$\pm$0.02\\
                                   
$\langle\log(L_K/L_{\odot})\rangle$   &   7.80$\pm$0.09 &   \underline{8.09$\pm$0.11} &   7.73$\pm$0.07 &   7.87$\pm$0.10 &   7.56$\pm$0.04\\
                                   
 $SB, ^m/\sq\arcsec $                 &  24.57$\pm$0.14 &  24.03$\pm$0.25 &  24.46$\pm$0.09 &  \underline{23.89$\pm$0.13} &  24.50$\pm$0.05\\
                                                            
$\langle\log(M_{HI}/M_{\odot})\rangle$ &  7.65$\pm$0.09  &  7.74$\pm$0.13  &  7.61$\pm$0.07  &  7.71$\pm$0.09  &  7.43$\pm$0.04\\

$\langle\log(M_{HI}/L_K)\rangle$       &  --0.21$\pm$0.08 &  --0.36$\pm$0.12 &  --0.18$\pm$0.06 &  --0.29$\pm$0.09 &  --0.22$\pm$0.03\\

$\langle\log(SFR_{H\alpha})\rangle$   &   \underline{--2.39$\pm$0.09} &  \underline{--1.87$\pm$0.18} &  --2.68$\pm$0.09 &  \underline{--1.95$\pm$0.09} &  --2.99$\pm$0.06\\
                                     
$\langle\log(SFR_{FUV})\rangle$       &   \underline{--2.24$\pm$0.11} &  \underline{--1.79$\pm$0.15} &  \underline{--2.39$\pm$0.08} &  \underline{--2.17$\pm$0.10} &  --2.69$\pm$0.05\\
                            
$\langle\log(SFR_{H\alpha}/L_K)\rangle$   &  \underline{--10.19$\pm$0.07} &  \underline{--9.96$\pm$0.09} & --10.41$\pm$0.06 &  \underline{--9.83$\pm$0.06} & --10.55$\pm$0.04\\
                                    
$\langle\log(SFR_{H\alpha}/SFR_{FUV})\rangle$      &   --0.22$\pm$0.06 &  --0.13$\pm$0.14 &  --0.33$\pm$0.05 &   \underline{0.14$\pm$0.04} &  --0.41$\pm$0.03\\
\hline
\end{tabular}
\end{table*}
 \clearpage                                                           
\input{Appendix.tex}

\clearpage
 
 \begin{figure*}
 \includegraphics[height=0.35\textwidth]{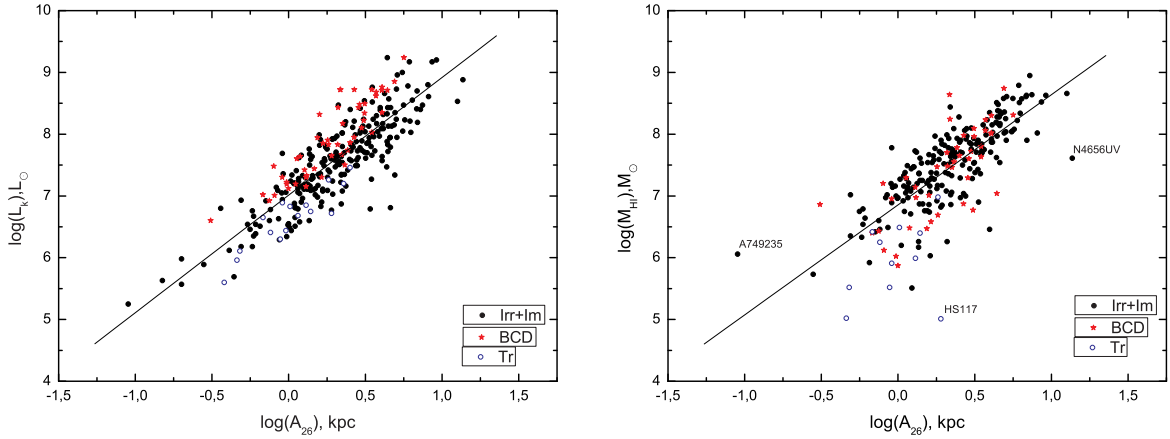}
 \caption{$K$-band luminosity (left panel) and hydrogen mass (right panel)              
       versus Holmberg linear diameter for late-type dwarf galaxies in 
         the LV. The galaxies of Irr and Im morphological types are indicated              
         by the solid circles, blue compact dwarfs are shown by asterisks,              
         and transition-type objects  are indicated by empty circles.}
         \end{figure*}
   
   \begin{figure*}
  \includegraphics[height=0.35\textwidth]{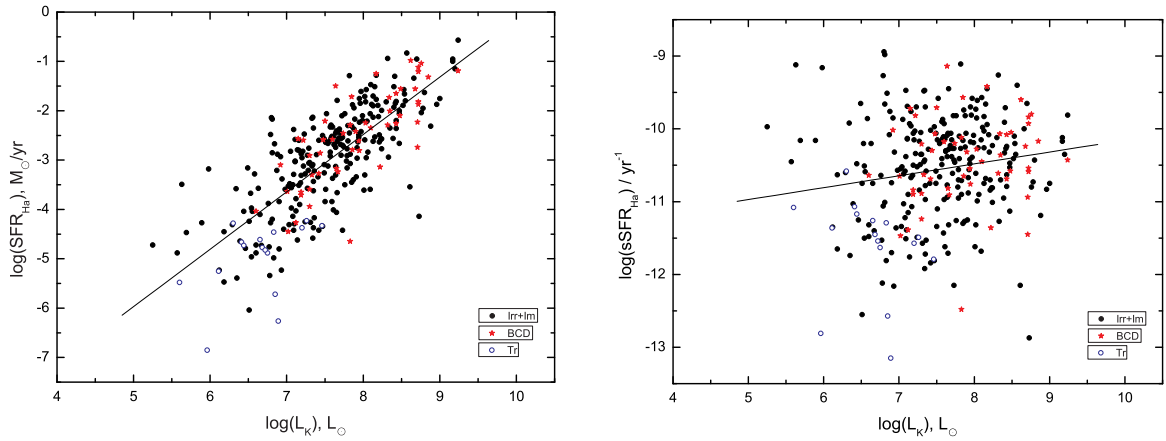}
  \caption{Integral star formation rate (left panel) and specific star
         formation rate (right panel) vs. $K$-band luminosity for the late-
         type galaxies. Objects of different types are indicated with the 
         same symbols as in Fig.1.}
         \end{figure*}

         \begin{figure*}
         \includegraphics[height=0.35\textwidth]{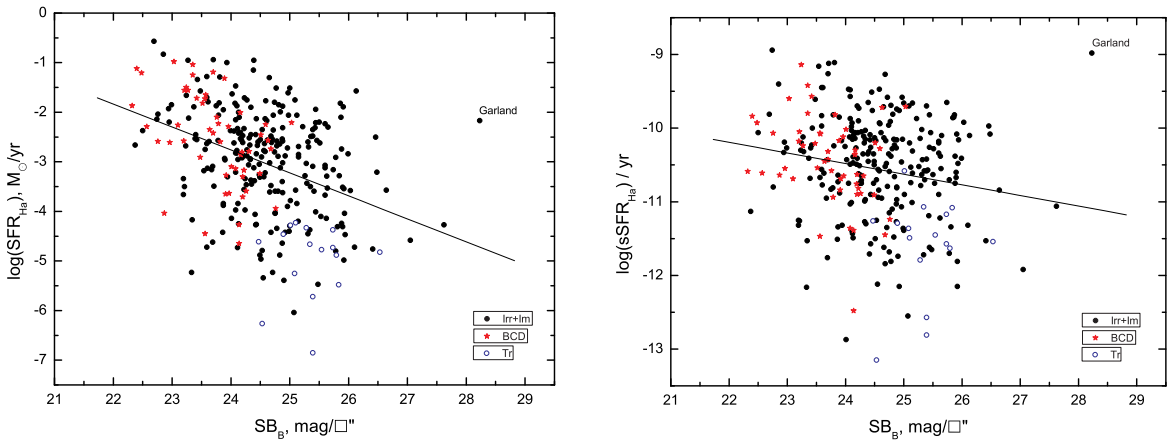}
  \caption{Integral star formation rate (left panel) and specific star
         formation rate (right panel) vs. the mean surface brightness
         in the $B$-band.
         Dwarf galaxies of different types are indicated with the same
         symbols as in Fig.1.}
         \end{figure*}

   \begin{figure*}
   \includegraphics[height=0.5\textwidth]{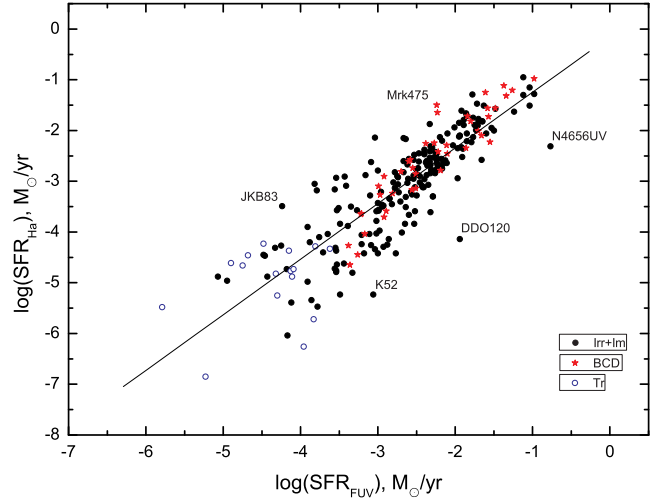}
  \caption{SFR from $H\alpha$ flux plotted against SFR from FUV flux for
         the late-type LV dwarfs.{The regression line has a slope of $1.09\pm0.04$
         as shown in Table 1}. }
         \end{figure*}

         \begin{figure*}
         \includegraphics[height=0.5\textwidth]{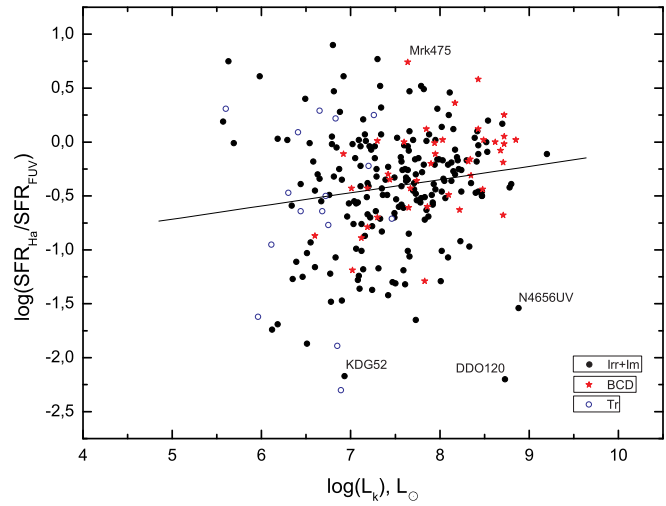}
  \caption{The ratio of  $H\alpha$-to-FUV SFRs as a function of K-band 
         luminosity. A typical error bar for K-band luminosity is $\pm0.2 dex$.}
   \end{figure*}
   
   \begin{figure*}
   \includegraphics[height=0.65\textwidth]{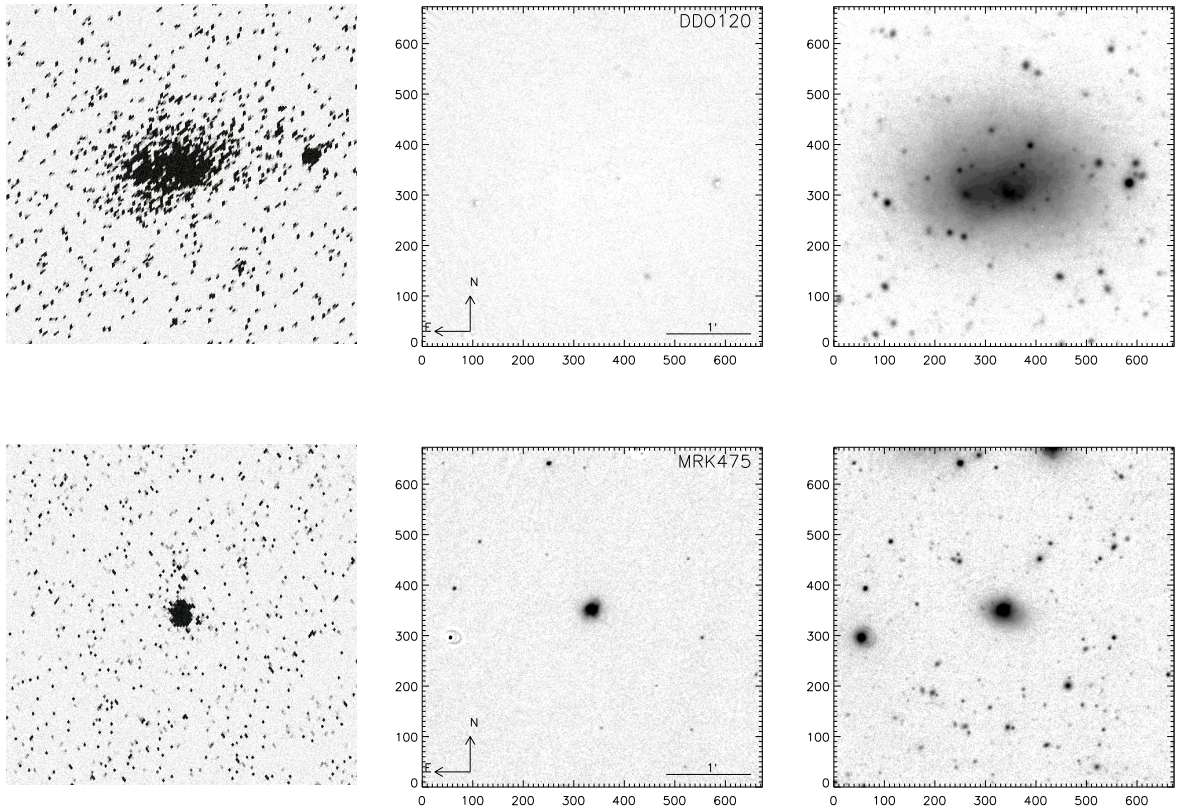}
  \caption{Images of two dwarf galaxies, DDO~120 and Mrk~475, staying in 
         different star-formation stages. The left-hand panel is taken
         from the GALEX survey. The middle panel and the right-hand panel correspond
         to continuum-subtracted $H\alpha$ line and red continuum, respectively. 
         The image size is $4\arcmin\times4\arcmin$.}
\end{figure*}

   \begin{figure*}
   \includegraphics[height=0.5\textwidth]{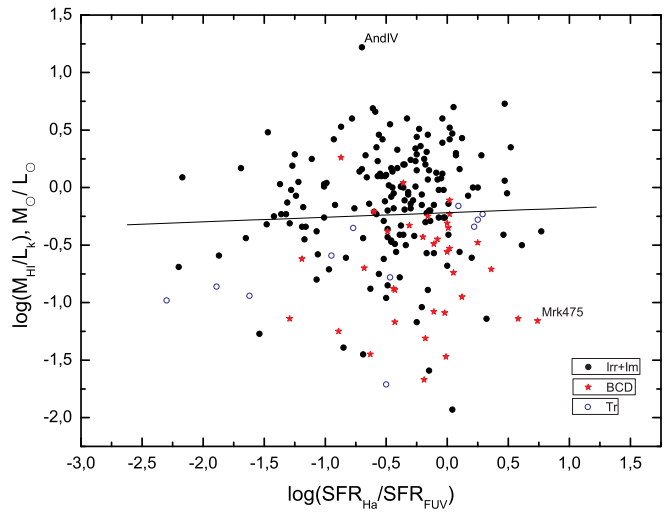}
  \caption{Hydrogen mass-to-K-luminosity ratio as a function of SFR($H\alpha$)-
         to-SFR(FUV) ratio.}
         \end{figure*}
     
\begin{figure*}
\includegraphics[height=1.25\textwidth]{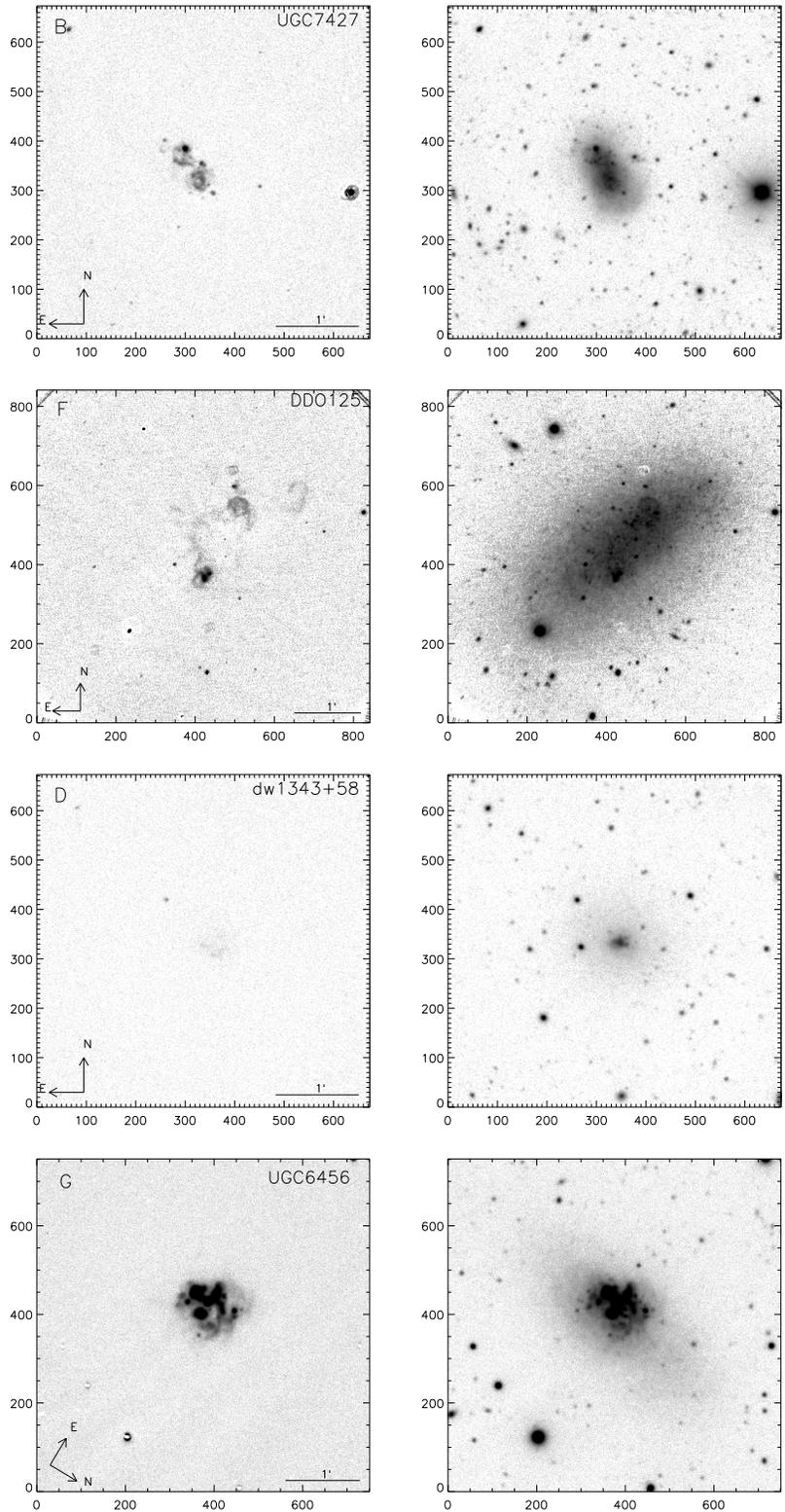}
  \caption{Four examples of galaxies with the specified properties: B -bubble,
         Ring-like structure similar to supernova remnants,
         F- filament, one-dimensional structure of various extension,
         D - diffuse emission, faint amorphous luminescence in the 
         presence or absence of compact $HII$-regions, and G - global burst,
         bright $H\alpha$ emission covering the main galaxy body. The right- 
         and left-hand images correspond to the red continuum and the 
         continuum-subtracted $H\alpha$  line. The image size is 
         $4\arcmin\times4\arcmin$.} 
   \end{figure*}
   
 \begin{figure*}
 \includegraphics[height=0.4\textwidth]{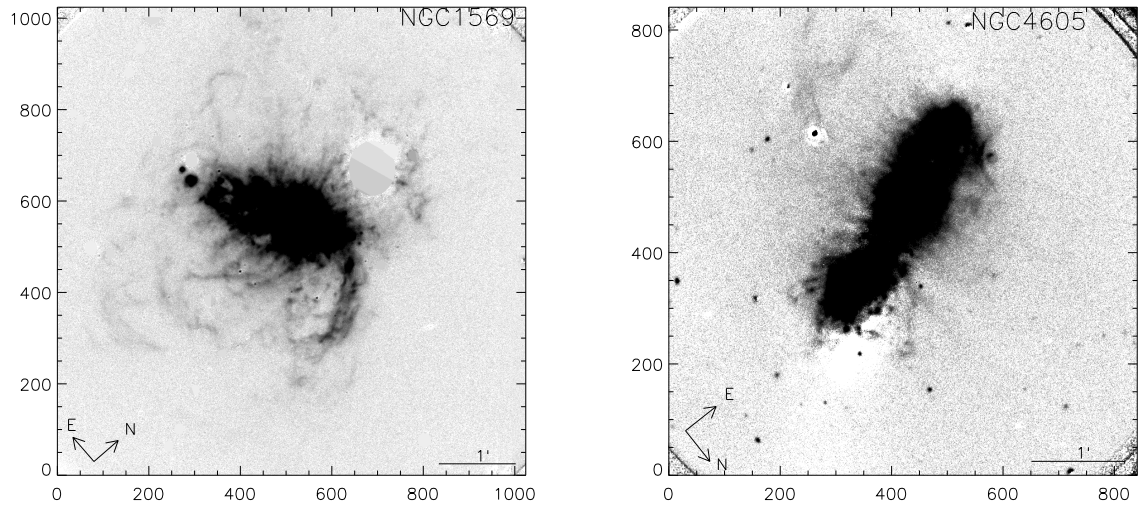}
  \caption{ Two Sm- galaxies in the Local Volume: NGC1569 and NGC4605 with rich 
         filament structures. The $H\alpha$ image size is $4\arcmin\times4\arcmin$.}
         \end{figure*}
         \end{document}

%% file: Appendix.tex
\renewcommand{\baselinestretch}{0.8}
\appendix{Appendix. Observational data on 309 LV dwarf galaxies. The full table is available online.}
 \small
\begin{center}
 \begin{tabular}{llcccccccccccc} \\ \hline
  
 \multicolumn{14}{l}{(1) --- galaxy name,}  \\                                                         
 \multicolumn{14}{l}{(2) ---morphological type,}\\                                                     
 \multicolumn{14}{l}{(3) --- distance in Mpc,}\\                                                        
 \multicolumn{14}{l}{(4) --- linear diameter in kpc measured within the Holmberg isophote,}\\           
 \multicolumn{14}{l}{(5) --- the average surface brightness in the $B$-band within the Holmberg        
       diameter, }\\                                                              
 \multicolumn{14}{l}{(6) --- logarithm of the integral luminosity of a galaxy in the $K$-band in      
       solar-luminosity units,}\\                                                
 \multicolumn{14}{l}{(7) --- logarithm of the neutral hydrogen mass in solar-mass units, }\\           
 \multicolumn{14}{l}{(8) --- logarithm of the integral star-formation rate ($M_{\odot}$/yr) determined    
       via the $H\alpha$ flux,}\\                                                  
 \multicolumn{14}{l}{(9) --- logarithm of the integral star-formation rate ($M_{\odot}$/yr) determined    
       via the $FUV$-band flux, }\\                                                
 \multicolumn{14}{l}{(10) --- number of compact emission knots,}\\                                     
 \multicolumn{14}{l}{(11) --- number of bubbles,}\\                                                    
 \multicolumn{14}{l}{(12) --- presence (1) or absence (0) of filament-like structures,}\\               
 \multicolumn{14}{l}{(13) --- presence (1) or absence (0) of a diffuse emission,}\\                    
 \multicolumn{14}{l}{(14) --- presence (1) or absence (0) of the global bright emission.}\\
 \hline
Name           &  T  & Dist&$A_{26}$ & $SB$ &  $L_K$ & $M_{HI}$& $SFR_{H\alpha}$& $SFR_{FUV}$&   N&  B& F& D& G\\
\hline
  (1)& (2)& (3)& (4)& (5) & (6)& (7)& (8)& (9)& (10)& (11)& (12)& (13)& (14)\\   \hline
UGC12894        &  Irr & 8.47 & 2.78&  25.22&  7.58 & 7.92 & $-$2.29 & $-$2.03  &  6 &  2&  0&  0&  0 \\
AGC748778       & Irr  & 6.22&  0.61 & 24.87 & 6.39 & 6.64 & $-$4.64 & $-$3.53  &  0 &  0&  0&  0&  0 \\
UGC00064        & Irr  & 9.60 & 4.87&  25.00&  8.15 & 8.58 & $-$1.51 & $-$1.63  &  7 &  0&  0&  0&  1 \\
ESO349$-$031      & Irr  & 3.21 & 1.15&  24.75&  7.12 & 7.13 & $-$4.03 & $-$3.02  &  0 &  0&  0&  1&  0 \\
UGC00288        & Irr  & 6.82 & 2.24&  24.20&  7.80 & 7.69 & $-$2.61 & $-$2.33  &  2 &  0&  0&  0&  0 \\
AndIV           & Irr  & 7.18 & 2.19&  25.60&  7.22 & 8.44 & $-$3.12 & $-$2.42  &  2 &  0&  0&  1&  0 \\
UGC00685        & BCD  & 4.81 & 3.49&  24.59&  8.03 & 7.80 & $-$2.25 & $-$2.27  &  5 &  0&  0&  0&  1 \\
KKH5            & Irr  & 4.27 & 1.38&  24.74&  7.16 & 6.87 & $-$2.80 &   $-$    &  3 &  0&  0&  0&  0 \\
UGC01056        & Im   & 10.2 & 3.12&  23.24&  8.54 & 7.93 & $-$1.66 & $-$1.86  &  2 &  0&  0&  0&  0 \\
KKH6            & Irr  & 5.18 & 1.79&  24.48&  7.30 & 6.92 & $-$3.05 & $-$3.82  &  0 &  0&  0&  1&  0 \\
KDG010          & Irr  & 7.87 & 3.49&  25.51&  7.66 & 7.72 & $-$2.62 & $-$3.09  &  2 &  0&  0&  1&  0 \\
KK13            & Irr  & 7.83 & 1.72&  24.19&  7.57 & 7.27 & $-$2.92 & $-$2.74  &  2 &  0&  0&  0&  0 \\
AGC112521       & Irr  & 6.08 & 1.10&  25.14&  6.81 & 6.75 & $-$4.20 & $-$3.88  &  1 &  0&  0&  0&  0 \\
DDO013          & Irr  & 9.04 & 12.6&  26.13&  8.53 & 8.66 & $-$1.57 & $-$1.48  & 14 &  2&  0&  0&  0 \\
UGC01171        & Irr  & 10.2 & 4.02&  24.75&  8.09 & 7.71 & $-$3.59 & $-$2.52  &  1 &  0&  0&  0&  0 \\
KK14            & Irr  & 7.05 & 1.67&  25.35&  7.08 & 7.54 & $-$3.50 & $-$2.94  &  2 &  0&  0&  0&  0 \\
KK15             &Irr  & 8.67 & 1.54&  25.41&  6.98 & 7.21 & $-$3.59 & $-$3.02   & 1  & 0 & 0 & 0 & 0 \\
Cas1             &Irr  & 5.27 & 8.61&  23.27&  9.17 & 8.52 & $-$0.95 &   $-$     &12  & 3 & 0 & 0 & 0 \\
UGC01561         &Im    &9.20  &3.51 & 23.37 & 8.52  &7.84  &$-$1.89  &$-$1.89   & 5  & 0 & 0 & 0 & 0 \\
KK17             &Irr   &5.01  &1.49 & 25.64 & 7.02  &6.76  &$-$3.53  &$-$3.51   & 1  & 0 & 0 & 0 & 0 \\
KK16             &Irr   &5.62  &2.21 & 25.29 & 7.35  &6.86  &$-$3.43  &$-$3.00   & 0  & 0 & 0 & 1 & 0 \\
$[$TT2009$]$25       &Tr    &9.95  &2.28 & 25.73 & 7.20  & $-$    &$-$4.37  &$-$4.15   & 0  & 0 & 0 & 0 & 0 \\
$[$TT2009$]$30       &Irr   &9.80  &3.40 & 27.62 & 6.79  & $-$    &$-$4.27  &$-$4.25   & 0  & 0 & 0 & 0 & 0 \\
KKH11            &Irr   &8.55  &6.39 & 24.74 & 8.55  &8.57  &$-$1.80  &  $-$     & 5  & 1 & 0 & 1 & 0 \\
DDO019           &Irr   &9.55  &6.27 & 25.86 & 8.03  &8.34  &$-$1.82  &$-$1.65   &12  & 1 & 0 & 0 & 0 \\
d0226+3325       &Irr   &9.55  &4.92 & 27.05 & 7.34  &8.02  &$-$4.58  &  $-$     & 0  & 0 & 0 & 0 & 0 \\
Halogas          &Irr   &9.55  &1.26 & 24.58 & 7.14  &7.14  &$-$2.79  &$-$3.00   & 1  & 0 & 0 & 0 & 1 \\
KKH12            &Irr   &3.48  &3.44 & 25.65 & 7.80  &7.36  &$-$2.30  &  $-$     & 2  & 0 & 0 & 1 & 0 \\
MB1              &Irr   &7.35  &5.51 & 24.27 & 9.00  &8.04  &$-$1.75  &  $-$     & 5  & 0 & 0 & 1 & 0 \\
UGC02172         &Irr   &9.30  &4.13 & 23.70 & 8.32  &7.87  &$-$0.94  &  $-$     & 1  & 0 & 1 & 0 & 1 \\
AGC122226        &BCD   &8.56  &2.32 & 25.03 & 7.50  &7.55  &$-$2.21  &  $-$     & 3  & 0 & 0 & 0 & 0 \\
Dw2              &Irr   &6.00  &5.11 & 23.75 & 8.96  &8.61  &$-$1.87  &  $-$     & 2  & 0 & 0 & 1 & 0 \\
MB3              &Irr   &5.47  &8.16 & 25.92 & 8.61  & $-$    &$-$3.54  &  $-$     & 0  & 0 & 0 & 0 & 0 \\
LVJ0300+2546     &Irr   &7.80  &1.05 & 23.69 & 7.34  &6.20  &$-$3.09  &$-$3.41   & 2  & 0 & 0 & 0 & 0 \\
KKH18            &Irr   &4.79  &1.16 & 24.08 & 7.28  &7.14  &$-$3.21  &$-$2.57   & 1  & 0 & 0 & 0 & 0 \\
UGC02684         &Irr   &6.49  &4.07 & 25.99 & 7.57  &7.92  &$-$2.84  &$-$2.26   & 5  & 0 & 0 & 0 & 0 \\
UGC02773         &Im    &7.52  &4.40 & 22.69 & 9.24  &8.51  &$-$0.57  &  $-$     &12  & 0 & 0 & 1 & 1 \\
UGC02905         &Irr   &5.83  &2.23 & 23.63 & 8.02  &7.56  &$-$2.59  &$-$2.13   & 2  & 0 & 0 & 1 & 0 \\
KK35             &Irr   &3.16  &3.91 & 24.97 & 7.97  & $-$    &$-$1.61  &$-$1.92   &18  & 2 & 1 & 1 & 1 \\
KKH22            &Tr    &3.28  &1.39 & 25.79 & 6.75  &6.40  &$-$4.88  &$-$4.11   & 0  & 0 & 0 & 0 & 0 \\
ESO483$-$013       &BCD   &7.40  &3.49 & 23.52 & 8.72  &7.63  &$-$1.82  &$-$1.80   & 1  & 0 & 0 & 0 & 0 \\
CamA             &Irr   &3.56  &4.61 & 25.78 & 7.79  &8.14  &$-$2.91  &$-$3.43   & 2  & 0 & 0 & 1 & 0 \\
NGC1592          &Irr   &9.10  &3.75 & 23.59 & 8.17  &7.99  &$-$1.28  &$-$0.98   & 3  & 0 & 1 & 1 & 1 \\
UGCA092          &Irr   &3.22  &3.81 & 23.48 & 8.04  &8.23  &$-$1.57  &  $-$     & 6  & 2 & 0 & 0 & 1 \\
CamB             &Irr   &3.50  &1.79 & 25.63 & 7.09  &7.12  &$-$3.29  &  $-$     & 1  & 2 & 0 & 1 & 0 \\
KKH30            &Tr    &4.50  &1.15 & 25.54 & 6.68  & $-$    &$-$4.77  &$-$4.13   & 0  & 0 & 0 & 0 & 0 \\
ESO553$-$046       &BCD   &6.70  &2.25 & 23.35 & 8.17  &7.46  &$-$1.25  &$-$1.61   & 1  & 0 & 0 & 0 & 1 \\
KK49             &BCD   &10.0  &5.64 & 23.70 & 9.24  &8.31  &$-$1.19  &  $-$     & 3  & 0 & 0 & 0 & 1 \\
KKH34            &Irr   &4.61  &1.55 & 24.96 & 7.17  &6.76  &$-$3.90  &$-$3.91   & 0  & 1 & 0 & 1 & 0 \\
UGCA127sat       &Irr   &8.50  &3.51 & 22.76 & 8.76  & $-$    &$-$2.04  &  $-$     & 1  & 0 & 0 & 1 & 0 \\
ESO490$-$017       &Irr   &6.34  &3.33 & 23.59 & 8.13  &7.90  &$-$1.70  &  $-$     & 2  & 0 & 0 & 1 & 0 \\
UGC03476         &Irr   &7.01  &3.59 & 24.81 & 7.96  &8.15  &$-$1.85  &  $-$     & 6  & 0 & 0 & 0 & 0 \\
ESO489$-$056       &Irr   &6.31  &1.37 & 23.42 & 7.68  &7.36  &$-$2.55  &  $-$     & 5  & 1 & 0 & 0 & 0 \\
KKSG9            &BCD   &10.0  &2.17 & 22.32 & 8.72  &8.64  &$-$1.87  &  $-$     & 1  & 0 & 0 & 0 & 0 \\
KKH37            &Irr   &3.44  &1.21 & 25.12 & 6.90  &6.71  &$-$3.84  &$-$3.49   & 1  & 0 & 0 & 1 & 0 \\
KKH38            &Irr   &10.2  &4.36 & 26.46 & 7.47  &8.19  &$-$2.50  &  $-$     & 8  & 1 & 0 & 0 & 0 \\
HIZSS003         &Irr   &1.67  &0.92 & 23.34 & 7.36  &7.14  &$-$4.17  &  $-$     & 1  & 0 & 0 & 0 & 0 \\
UGC03817         &Irr   &8.55  &4.62 & 25.50 & 8.43  &8.17  &$-$2.20  &$-$2.17   &13  & 2 & 0 & 0 & 0 \\
UGC03755         &Irr   &7.69  &4.43 & 23.80 & 8.43  &8.28  &$-$1.81  &$-$1.7    & 8  & 0 & 1 & 1 & 0 \\
AGC174585        &Irr   &7.73  &0.91 & 24.40 & 6.94  &6.89  &$-$3.38  &  $-$     & 0  & 0 & 0 & 1 & 0 \\
 \end{tabular}

 \begin{tabular}{llcccccccccccc} \\ \hline
Name           &  T  & Dist&$A_{26}$ & $SB$ &  $L_K$ & $M_{HI}$& $SFR_{H\alpha}$& $SFR_{FUV}$&   N&  B& F& D& G\\
\hline
  (1)& (2)& (3)& (4)& (5) & (6)& (7)& (8)& (9)& (10)& (11)& (12)& (13)& (14)\\
\hline
KK65             &Irr   &7.98  &2.62 & 24.10 & 8.11  &7.70  &$-$1.87  &$-$2.33   & 2  & 0 & 0 & 0 & 0 \\
KKH40            &Irr   &9.25  &1.86 & 24.16 & 7.65  &7.72  &$-$2.41  &$-$2.33   & 6  & 0 & 0 & 1 & 0 \\
AGC174605        &Irr   &9.59  &0.96 & 24.22 & 7.06  &7.16  &$-$3.56  &$-$2.99   & 1  & 0 & 0 & 0 & 0 \\
HIPASSJ0801$-$21   &Irr   &10.4  &1.74 & 23.67 & 7.79  &7.97  &$-$3.16  &  $-$     & 0  & 0 & 0 & 0 & 0 \\
KUG0821+321      &BCD   &13.5  &2.09 & 23.96 & 7.83  &7.70  &$-$2.29  &  $-$     & 3  & 1 & 0 & 0 & 0 \\
KDG052           &Irr   &3.42  &0.49 & 23.33 & 6.93  &7.02  &$-$5.23  &$-$3.06   & 0  & 0 & 0 & 0 & 0 \\
HolmII           &Im    &3.47  &9.21 & 24.38 & 9.20  &8.63  &$-$1.15  &$-$1.04   &15  & 4 & 0 & 1 & 1 \\
DDO053           &Irr   &3.68  &2.20 & 24.58 & 7.34  &7.64  &$-$2.13  &$-$2.20   & 5  & 2 & 1 & 1 & 1 \\
UGC04483         &Irr   &3.58  &1.39 & 24.05 & 7.09  &7.61  &$-$2.38  &$-$2.40   & 2  & 2 & 0 & 0 & 1 \\
AGC182595        &BCD   &8.47  &0.91 & 23.48 & 7.30  &6.95  &$-$2.91  &$-$2.92   & 1  & 0 & 0 & 1 & 0 \\
KK69             &Irr   &9.16  &3.47 & 26.64 & 7.27  &7.66  &$-$3.57  &$-$3.53   & 0  & 0 & 0 & 0 & 0 \\
LV0853+3318      &Irr   &9.82  &1.01 & 25.74 & 6.49  & $-$    &$-$3.16  &$-$3.56   & 0  & 0 & 0 & 1 & 0 \\
KKH46            &Irr   &6.70  &1.22 & 24.46 & 7.16  &7.44  &$-$2.60  &$-$2.67   & 5  & 0 & 0 & 0 & 1 \\
LVJ0913+1937     &Irr   &8.90  &1.41 & 24.52 & 7.27  &7.07  &$-$2.90  &$-$2.75   & 2  & 0 & 0 & 0 & 0 \\
HIPASSJ0916$-$23b  &Irr   &10.3  &2.11 & 22.37 & 8.47  &7.51  &$-$2.66  &$-$2.16   & 1  & 0 & 0 & 0 & 0 \\
AGC198508        &Irr   &10.4  &1.16 & 24.20 & 7.23  &6.97  &$-$2.47  &$-$2.40   & 2  & 1 & 0 & 0 & 0 \\
ESO565$-$003       &Irr   &10.7  &2.35 & 23.28 & 8.20  &7.63  &$-$2.12  &$-$1.95   & 2  & 0 & 0 & 0 & 0 \\
UGC04998         &BCD   &8.24  &4.41 & 24.68 & 8.71  &7.04  &$-$2.74  &$-$2.55   & 6  & 0 & 0 & 0 & 0 \\
KUG0937+480      &Im    &11.8  &2.84 & 25.15 & 7.62  & $-$    &$-$2.69  &$-$2.36   & 3  & 0 & 0 & 0 & 0 \\
KDG058           &Irr   &10.0  &2.01 & 24.78 & 7.47  & $-$    &$-$4.34  &$-$3.04   & 0  & 0 & 0 & 0 & 0 \\
HolmI            &Irr   &4.02  &5.54 & 25.44 & 8.05  &8.05  &$-$2.21  &$-$1.85   &12  & 2 & 0 & 0 & 0 \\
UGC05186         &Irr   &8.30  &3.18 & 25.44 & 7.61  &7.38  &$-$4.01  &$-$2.69   & 1  & 0 & 0 & 0 & 0 \\
NGC2903$-$HI$-$1     &Irr   &8.90  &0.71 & 23.89 & 6.92  &6.42  &$-$2.93  &$-$3.54   & 2  & 0 & 0 & 0 & 0 \\
KDG056           &Irr   &8.90  &1.81 & 24.68 & 7.42  &7.17  &$-$4.42  &$-$3.00   & 0  & 0 & 0 & 0 & 0 \\
$[$CKT2009$]$d0926+70&Tr    &3.40  &0.48 & 25.08 & 6.11  &5.52  &$-$5.25  &$-$4.30   & 0  & 0&  0&  0 & 0 \\
$[$CKT2009$]$d0939+71&Tr    &3.65  &0.38 & 25.83 & 5.60  & $-$    &$-$5.48  &$-$5.79   & 0  & 0&  0&  0 & 0 \\
LVJ0939$-$2507     &Irr   &8.32  &1.58 & 22.95 & 7.99  &7.25  &$-$2.20  &  $-$     & 3  & 1 & 0 & 0 & 0 \\
BK3N             &Irr  &4.17 &0.41& 24.71& 6.12  & $-$    &$-$5.23  &$-$3.49   & 0  & 0 & 0 & 1 & 0 \\
KKSG15           &Irr   &9.70  &3.56 & 24.10 & 8.24  &7.77  &$-$2.77  &$-$2.31   & 0  & 0 & 0 & 1 & 0 \\
LVJ0956$-$0929     &BCD   &9.38  &2.67 & 24.19 & 7.95  &6.87  &$-$2.81  &$-$2.70   & 1  & 0 & 0 & 0 & 0 \\
KDG61em          &Irr   &3.70  &0.20 & 23.54 & 5.98  & $-$    &$-$3.18  &$-$3.79   & 1  & 0 & 0 & 0 & 1 \\
ClumpI           &Irr   &3.60  &0.20 & 24.49 & 5.57  & $-$    &$-$4.88  &$-$5.07   & 0  & 0 & 0 & 1 & 0 \\
A0952+69         &Irr   &3.93  &2.20 & 26.48 & 6.87  &7.02  &$-$3.21  &$-$2.96   & 3  & 0 & 0 & 1 & 0 \\
$[$CKT2009$]$d0958+66&BCD   &3.82  &1.00 & 24.14 & 7.12  &5.87  &$-$4.27  &$-$3.38   & 1  & 0& 0& 0& 0    \\
$[$CKT2009$]$d0959+68&Irr   &4.27  &1.04 & 25.93 & 6.44  & $-$    &$-$4.04  &$-$3.65   & 3  & 0& 0& 1& 0    \\
LVJ1000+5022     &Im    &11.8  &1.99& 24.50& 7.58    & $-$    &$-$2.55 &  $-$      & 2  & 0 & 0 & 0 & 0 \\
LVJ1000+3032     &Irr  &7.10 &0.58& 23.87& 6.76 &6.33 &$-$3.88 &$-$3.39   & 1  & 0 & 0 & 0 & 0 \\
MCG$-$01$-$26$-$009    &Irr   &9.70  &3.24 & 24.88 & 7.85  &7.71  &$-$2.31  &$-$2.32   & 1  & 0 & 0 & 0 & 0 \\
KKSG17           &Irr   &6.92  &3.32 & 24.87 & 7.87  &7.43  &$-$3.40  &$-$2.71   & 0  & 0 & 0 & 1 & 0 \\
GARLAND          &Irr   &3.82  &4.61 & 28.23 & 6.81  &7.54  &$-$2.17  &$-$2.64   &18  & 1 & 0 & 1 & 1 \\
UGC05423         &Irr   &8.87  &2.93 & 22.99 & 8.42  &7.86  &$-$1.85  &$-$1.95   & 4  & 0 & 0 & 0 & 1 \\
UGC05497         &BCD   &3.73  &0.97 & 23.91 & 7.19  &6.02  &$-$3.65  &$-$3.22   & 0  & 1 & 0 & 1 & 0 \\
LVJ1017+2922     &BCD   &6.79  &1.31 & 24.20 & 7.33  &6.97  &$-$3.30  &  $-$     & 1  & 0 & 0 & 0 & 0 \\
HS117            &Tr    &3.96  &1.90 & 26.53 & 6.72  &5.01  &$-$4.82  &$-$4.32   & 1  & 0 & 0 & 0 & 0 \\
LV1021+0054      &Irr   &8.59  &1.09 & 24.10 & 7.21  &7.33  &$-$2.82  &$-$2.78   & 3  & 0 & 0 & 0 & 1 \\
LeoP             &Irr   &1.73  &0.28 & 24.42 & 5.89  &5.73  &$-$4.27  &  $-$     & 1  & 0 & 0 & 1 & 0 \\
AGC731457        &Irr   &10.5  &1.17 & 23.23 & 7.62  &7.21  &$-$2.68  &$-$2.32   & 3  & 0 & 0 & 0 & 0 \\
LVJ1030+0607     &Irr   &7.80  &1.61 & 24.58 & 7.36  &7.16  &$-$3.22  &$-$2.73   & 2  & 0 & 0 & 0 & 0 \\
DDO082           &Im    &3.93  &3.95 & 24.62 & 8.39  &6.46  &$-$2.56  &$-$2.60   & 4  & 1 & 0 & 1 & 1 \\
AGC749315        &BCD   &9.46  &0.31 & 22.87 & 6.60  &6.86  &$-$4.04  &$-$3.17   & 1  & 0 & 0 & 0 & 0 \\
$[$CKT2009$]$d1028+70&BCD   &3.84  &0.81 & 23.97 & 7.01  &6.12  &$-$3.64  &$-$3.21   & 0  & 1& 0& 0& 0 \\
KUG1033+366B     &Im    &11.9  &2.06 & 24.42 & 7.64  & $-$    &$-$2.45  &$-$2.41   & 4  & 0 & 0 & 0 & 0 \\
LeG06            &Irr   &10.4  &1.94 &25.81  & 7.03  &6.85  &$-$4.31  &$-$3.55   & 0  & 0 & 0  & 0 & 0\\
DDO087           &Irr   &8.51  &3.95 & 24.72 & 8.43  &8.35  &$-$2.36  &$-$1.90   & 8  & 0 & 0 & 0 & 0 \\
LVJ1052+3628     &Irr   &9.20  &1.18 & 24.68 & 7.04  & $-$    &$-$3.48  &$-$2.93   & 2  & 0 & 0 & 0 & 0 \\
KDG073           &Irr   &3.91  &1.36 & 26.10 & 6.60  &6.26  &$-$4.72  &$-$3.56   & 0  & 0 & 0 & 0 & 0 \\
UGC06145         &Irr   &10.7  &4.42 & 25.60 & 7.83  &7.97  &$-$3.07  &$-$2.35   & 1  & 0 & 0 & 0 & 0 \\
UGC06456         &Irr   &4.63  &2.01 & 23.61 & 7.66  &7.82  &$-$1.79  &$-$1.89   & 7  & 0 & 0 & 0 & 1 \\
KK109            &Irr   &4.51  &0.59 & 24.90 & 6.35  &6.54  &$-$5.39  &$-$4.12   & 0  & 0 & 0 & 0 & 0 \\
KKH73            &Tr    &9.00  &1.82 & 25.10 & 7.26  &6.98  &$-$4.23  &$-$4.48   & 0  & 0 & 0 & 0 & 0 \\
LVJ1157+5638     &Irr   &8.75  &1.23 & 24.08 & 7.33  & $-$    &$-$2.15  &$-$2.67   & 1  & 0 & 0 & 0 & 1 \\
UGC07242         &Irr   &5.45  &1.92 & 23.56 & 7.92  &7.69  &$-$2.75  &$-$2.17   & 3  & 0 & 0 & 0 & 0 \\
MCG+09$-$20$-$131    &Irr   &4.61  &1.37 & 23.87 & 7.50  &7.37  &$-$3.84  &$-$2.53   & 2  & 0 & 0 & 0 & 0 \\
UGC07298         &Irr   &4.19  &1.03 & 24.13 & 7.24  &7.27  &$-$4.25  &$-$2.88   & 0  & 0 & 0 & 1 & 0 \\
LVJ1217+4703     &Tr    &7.66  &0.68 & 24.47 & 6.65  &6.42  &$-$4.61  &$-$4.90   & 0  & 0 & 0 & 0 & 0 \\
KKH78            &Irr   &4.70  &0.61 & 24.51 & 6.54  & $-$    &$-$4.96  &$-$4.95   & 0  & 0 & 0 & 0 & 0 \\
KUG1218+387      &BCD   &7.35  &2.52 & 24.31 & 7.86  &7.65  &$-$2.79  &$-$2.19   & 2  & 1 & 0 & 0 & 0 \\
DDO120           &Im    &7.28  &5.04 & 24.01 & 8.73  &8.04  &$-$4.14  &$-$1.94   & 0  & 0 & 0 & 0 & 0 \\
UGC7427          &Irr   &9.70  &3.00 & 24.61 & 7.89  &7.76  &$-$2.17  &  $-$     & 3  & 2 & 0 & 0 & 0 \\
KK138            &Tr    &6.30  &0.76 & 25.34 & 6.41  &6.25  &$-$4.66  &$-$4.75   & 0  & 0 & 0 & 0 & 0 \\
GR34             &Irr   &9.29  &2.55 & 24.35 & 7.85  &6.97  &$-$3.45  &$-$2.82   & 1  & 0 & 0 & 0 & 0 \\
KK141            &Irr   &7.78  &0.90 & 24.08 & 7.05  &7.20  &$-$2.82  &$-$2.67   & 1  & 0 & 0 & 1 & 0 \\
KK144            &Irr   &6.00  &1.93 & 25.26 & 7.24  &7.84  &$-$3.28  &$-$2.50   & 2  & 1 & 0 & 0 & 0 \\
UGCA281          &Irr   &5.70  &1.74 & 23.81 & 7.82  &7.77  &$-$1.29  &$-$1.78   & 2  & 1 & 0 & 0 & 1 \\
SBS1224+533      &BCD   &6.28  &1.47 & 24.17 & 7.44  & $-$    &$-$2.86  &$-$2.51   & 1  & 0 & 0 & 0 & 1 \\
DDO125           &Im    &2.61  &2.94 & 24.43 & 8.06  &7.44  &$-$2.84  &$-$2.32   & 2  & 2 & 1 & 0 & 0 \\
UGC07584         &BCD   &9.20  &2.43 & 24.51 & 7.74  &7.78  &$-$2.46  &$-$2.10   & 3  & 0 & 0 & 1 & 0 \\
 \end{tabular}

 \begin{tabular}{llcccccccccccc} \\ \hline
Name           &  T  & Dist&$A_{26}$ & $SB$ &  $L_K$ & $M_{HI}$& $SFR_{H\alpha}$& $SFR_{FUV}$&   N&  B& F& D& G\\
\hline
  (1)& (2)& (3)& (4)& (5) & (6)& (7)& (8)& (9)& (10)& (11)& (12)& (13)& (14)\\
\hline
KKH80            &Irr   &5.83  &1.45 & 25.00 & 7.10  &6.87  &$-$4.29  &$-$2.93   & 0  & 0 & 0 & 0 & 0 \\
DDO127           &Irr   &4.72  &2.20 & 24.56 & 7.64  &7.67  &$-$3.30  &$-$2.29   & 4  & 1 & 0 & 0 & 0 \\
UGC07605         &Irr   &4.74  &2.04 & 24.18 & 7.71  &7.38  &$-$2.56  &$-$2.18   & 2  & 0 & 0 & 1 & 0 \\
KK151            &Im    &8.20  &2.63 & 24.56 & 7.79  &7.50  &$-$2.77  &$-$2.25   & 3  & 0 & 0 & 1 & 0 \\
UGC07639         &Im    &7.14  &4.38 & 24.33 & 8.33  &7.62  &$-$2.94  &$-$1.97   & 2  & 0 & 0 & 1 & 0 \\
KK149            &Irr   &8.51  &2.32 & 23.39 & 8.15  &7.73  &$-$2.53  &$-$2.26   & 1  & 0 & 0 & 1 & 0 \\
DDO131           &Irr   &8.10  &2.72 & 24.14 & 7.99  &8.06  &$-$2.10  &$-$1.94   & 6  & 3 & 0 & 0 & 0 \\
NGC4509          &BCD   &10.1  &2.68 & 22.48 & 8.72  &7.98  &$-$1.21  &$-$1.26   & 5  & 1 & 0 & 0 & 1 \\
KK152            &Irr   &9.77  &3.00 & 24.96 & 7.74  &7.84  &$-$2.72  &$-$2.18   & 1  & 0 & 0 & 1 & 0 \\
IC3583           &Im    &9.51  &8.10 & 24.02 & 8.80  &8.02  &$-$1.63  &$-$1.24   & 5  & 0 & 0 & 1 & 0 \\
UGCA292          &Irr   &3.85  &1.15 & 24.68 & 6.79  &7.49  &$-$2.48  &$-$2.53   & 6  & 0 & 0 & 0 & 1 \\
BTS146           &Irr   &8.50  &0.83 & 23.69 & 7.14  &6.97  &$-$4.62  &$-$3.44   & 1  & 0 & 0 & 0 & 0 \\
KDG192           &Irr   &7.83  &2.43 & 25.31 & 7.42  &7.93  &$-$2.60  &$-$2.37   & 3  & 1 & 0 & 1 & 0 \\
LVJ1243+4127     &Irr   &4.81  &1.90 & 26.41 & 6.77  &6.82  &$-$4.76  &$-$3.54   & 0  & 0 & 0 & 0 & 0 \\
KK160            &Irr   &4.33  &0.75 & 24.83 & 6.60  &6.60  &$-$4.88  &$-$4.43   & 0  & 0 & 0 & 0 & 0 \\
UGCA294          &BCD   &9.90  &3.13 & 23.56 & 8.34  &8.09  &$-$1.73  &$-$1.57   & 4  & 0 & 0 & 0 & 1 \\
DDO147           &Irr   &3.01  &1.27 & 24.49 & 7.19  &7.47  &$-$3.31  &$-$2.64   & 3  & 2 & 0 & 0 & 0 \\
IC3840           &Irr   &5.97  &1.67 & 25.28 & 7.11  &7.28  &$-$3.03  &$-$2.62   & 1  & 1 & 0 & 0 & 0 \\
KDG215           &Irr   &4.83  &1.64 & 25.78 & 6.90  &7.38  &$-$4.80  &$-$3.33   & 0  & 0 & 0 & 0 & 0 \\
UGC08215         &Irr   &4.57  &1.13 & 24.26 & 7.18  &7.28  &$-$3.46  &$-$2.95   & 1  & 0 & 0 & 0 & 0 \\
DDO165           &Im    &4.83  &5.71 & 24.88 & 8.23  &8.08  &$-$2.58  &$-$1.66   & 4  & 0 & 0 & 1 & 0 \\
UGC08245         &Im    &4.72  &2.92 & 24.88 & 7.78  & $-$    &$-$3.02  &$-$2.46   & 2  & 0 & 0 & 0 & 0 \\
KK191            &Tr    &8.28  &1.02 & 24.89 & 6.83  &6.49  &$-$4.46  &$-$4.68   & 0  & 0 & 0 & 0 & 0 \\
CGCG189$-$050      &BCD   &3.93  &0.97 & 23.82 & 7.22  & $-$    &$-$2.60  &  $-$     & 3  & 0 & 0 & 0 & 0 \\
DDO169NW         &Irr   &4.33  &0.97 & 26.03 & 6.34  &7.00  &$-$3.58  &$-$2.99   & 2  & 0 & 0 & 0 & 0 \\
UGC08508         &Irr   &2.67  &1.65 & 24.32 & 7.54  &7.29  &$-$2.66  &  $-$     & 6  & 0 & 0 & 0 & 1 \\
MCG+08$-$25$-$028    &Irr   &8.40  &2.31 & 24.34 & 7.7   & $-$    &$-$2.45  &$-$2.47   & 3  & 0 & 0 & 0 & 1 \\
UGC08638         &Irr   &4.29  &2.06 & 24.13 & 7.97  &7.08  &$-$2.42  &$-$2.26   & 2  & 1 & 1 & 0 & 1 \\
DDO181           &Irr   &3.10  &2.13 & 24.69 & 7.56  &7.36  &$-$2.61  &$-$2.47   & 3  & 1 & 0 & 0 & 0 \\
KKH86            &Irr   &2.61  &0.65 & 25.07 & 6.51  &5.92  &$-$6.04  &$-$4.17   & 0  & 0 & 0 & 0 & 0 \\
KK230            &Irr   &2.21  &0.49 & 25.48 & 6.18  &6.35  &$-$5.47  &$-$3.78   & 0  & 0 & 0 & 0 & 0 \\
KKH87            &Irr   &8.87  &2.81 & 24.87 & 7.73  &7.97  &$-$2.54  &$-$2.24   & 5  & 0 & 0 & 0 & 0 \\
MRK0475          &BCD   &9.20  &1.19 & 23.24 & 7.64  &6.48  &$-$1.50  &$-$2.24   & 1  & 0 & 0 & 0 & 1 \\
ESO222$-$010       &Irr   &5.80  &0.91 & 22.50 & 7.69  &7.78  &$-$2.37  &  $-$     & 3  & 0 & 0 & 1 & 1 \\
PGC051659        &Irr   &3.61  &2.13 & 24.80 & 7.51  &7.78  &$-$3.59  &  $-$     & 1  & 0 & 0 & 1 & 0 \\
DDO190           &Im    &2.83  &1.99 & 23.59 & 7.93  &7.52  &$-$2.63  &$-$2.24   & 5  & 1 & 1 & 1 & 0 \\
ESO272$-$025       &Irr   &5.30  &2.76 & 23.99 & 7.88  &7.00  &$-$2.37  &  $-$     & 3  & 1 & 1 & 1 & 1 \\
UGC09660         &BCD   &10.7  &3.73 & 23.27 & 8.68  &8.23  &$-$1.56  &$-$1.48   & 6  & 0 & 0 & 0 & 1 \\
IC4662           &BCD   &2.55  &2.18 & 22.40 & 8.72  &8.24  &$-$1.12  &$-$1.37   &10  & 0 & 1 & 1 & 1 \\
KDG235           &Irr   &10.6  &4.07 & 25.88 & 7.64  &7.87  &$-$2.74  &$-$2.48   & 2  & 0 & 0 & 1 & 0 \\
UGC11411         &BCD   &6.58  &1.68 & 23.42 & 7.85  & $-$    &$-$1.72  &$-$1.84   & 3  & 0 & 0 & 1 & 1 \\
NGC6789          &BCD   &3.55  &1.54 & 22.96 & 7.94  &6.47  &$-$2.61  &$-$2.60   & 3  & 0 & 0 & 0 & 1 \\
SagdIrr          & Irr  &1.08  &1.00 & 24.75 & 6.55  &6.97  &$-$4.04  &$-$3.11   & 2  & 0 & 0 & 1 & 0 \\
KKR56            &Irr   &7.73  &3.08 & 23.78 & 8.24  & $-$    &$-$2.57  &  $-$     & 4  & 0 & 0 & 1 & 0 \\
KKR55            &Irr   &7.73  &2.52 & 22.94 & 8.40  &7.87  &$-$1.93  &  $-$     & 3  & 0 & 0 & 0 & 1 \\
KK252            &Irr   &7.73  &3.05 & 24.05 & 8.13  &7.28  &$-$2.62  &  $-$     & 4  & 0 & 0 & 0 & 0 \\
KK251            &Irr   &7.73  &4.57 & 25.39 & 7.94  &8.28  &$-$2.48  &  $-$     & 3  & 0 & 0 & 1 & 0 \\
UGC11583         &Irr   &7.73  &6.18 & 25.36 & 8.21  &8.51  &$-$2.49  &  $-$     & 7  & 0 & 0 & 1 & 0 \\
KKR60            &Irr   &7.73  &4.22 & 23.43 & 8.66  & $-$    &$-$1.34  &  $-$     & 4  & 0 & 0 & 1 & 0 \\
IC5152           &BCD   &1.96  &4.05 & 23.79 & 8.71  &8.01  &$-$2.23  &$-$1.55   &15  & 1 & 0 & 1 & 0 \\
UGCA438          &Irr   &2.22  &1.39 & 24.12 & 7.59  &7.25  &$-$3.84  &$-$2.65   & 3  & 0 & 0 & 1 & 0 \\
KKH98            &Irr   &2.58  &0.86 & 25.10 & 6.63  &6.49  &$-$3.57  &$-$3.27   & 1  & 0 & 0 & 0 & 0 \\
ESO149$-$003       &Irr   &7.01  &5.08 & 25.60 & 7.95  &7.90  &$-$2.23  &$-$1.79   & 1  & 0 & 0 & 1 & 0 \\
PiscesA          &Irr   &5.65  &0.99 & 25.60 & 6.53  &6.96  &$-$3.67  &   $-$    & 1  & 0 & 0 & 1 & 0 \\
PiscesB          &Irr   &8.91  &1.26 & 24.66 & 7.11  &7.47  &$-$3.02  &$-$2.80   & 4  & 0 & 0 & 0 & 0 \\
JKB129           &Irr   &8.90  &1.85 & 24.80 & 7.39  &7.59  &$-$2.89  &$-$2.54   & 2  & 0 & 0 & 1 & 0 \\
AGC112454        &Im    &10.2  &2.31 & 25.39 & 7.35  &8.09  &$-$2.35  &  $-$     & 3  & 0 & 0 & 1 & 0 \\
AGC114027        &Irr   &9.90  &1.11 & 25.15 & 6.81  &7.38  &$-$2.85  &  $-$     & 3  & 0 & 0 & 0 & 0 \\
AGC112503        &BCD   &10.2  &1.30 & 24.63 & 7.15  &7.14  &$-$2.57  &  $-$     & 2  & 0 & 0 & 0 & 0 \\
N672dwB          &Irr   &7.20  &0.44 & 25.94 & 5.69  & $-$    &$-$4.47  &$-$4.46   & 0  & 0 & 0 & 0 & 0 \\
N672dwA          &Irr   &7.20  &0.58 & 25.31 & 6.18  & $-$    &$-$4.45  &$-$4.48   & 0  & 0 & 0 & 0 & 0 \\
AGC123352        &Irr   &9.46  &1.09 & 24.92 & 6.88  &7.16  &$-$2.88  &$-$3.16   & 1  & 0 & 0 & 0 & 1 \\
AGC124056        &Irr   &7.80  &0.59 & 24.11 & 6.67  &6.79  &$-$4.73  &$-$4.18   & 0  & 0 & 0 & 0 & 0 \\
N1156dw1         &Irr   &7.80  &1.10 & 25.53 & 6.65  & $-$    &$-$4.10  &$-$3.76   & 0  & 0 & 0 & 0 & 0 \\
N1156dw2         &Irr   &7.80  &1.10 & 25.88 & 6.51  & $-$    &$-$4.24  &$-$3.21   & 0  & 0 & 0 & 0 & 0 \\
GALFA$-$Dw4        &Irr   &7.22  &1.78 & 23.79 & 7.76  &7.37  &$-$2.06  &  $-$     & 2  & 0 & 0 & 0 & 0 \\
LV J0831+4104    &BCD   &7.90  &0.68 & 23.56 & 7.02  &6.40  &$-$4.45  &$-$3.26   & 0  & 0 & 0 & 1 & 0 \\
LV J0843+4025    &BCD   &7.80  &0.75 & 24.01 & 6.92  &6.43  &$-$3.10  &$-$2.99   & 2  & 0 & 0 & 0 & 1 \\
JKB83            &Irr   &3.70  &0.15 & 23.71 & 5.63  & $-$    &$-$3.49  &$-$4.24   & 1  & 1 & 0 & 0 & 1 \\
UGC05571         &Im    &8.24  &2.80 & 25.46 & 7.49  &8.09  &$-$2.57  &$-$2.24   & 9  & 0 & 0 & 0 & 0 \\
LV J1028+4240    &Irr   &10.3  &1.43 & 24.39 & 7.33  & $-$    &$-$3.12  &$-$2.41   & 2  & 0 & 0 & 0 & 0 \\
PGC2277751       &BCD   &9.73  &1.30 & 24.26 & 7.30  & $-$    &$-$3.59  &$-$2.89   & 1  & 0 & 0 & 0 & 0 \\
N3344dw1         &Irr   &9.82  &0.87 & 25.90 & 6.29  & $-$    &$-$4.31  &$-$4.33   & 0  & 0 & 0 & 0 & 0 \\
LV J1052+3639    &Irr   &9.20  &4.38 & 25.44 & 7.88  & $-$    &$-$1.94  &$-$2.01   & 5  & 0 & 0 & 1 & 1 \\
HS1053+3624      &Irr   &9.20  &1.32 & 24.78 & 7.11  &7.71  &$-$2.58  &$-$2.54   & 4  & 0 & 0 & 0 & 1 \\
VV747            &Irr   &9.20  &2.92 & 24.23 & 8.02  & $-$    &$-$1.71  &$-$1.85   & 2  & 1 & 0 & 1 & 1 \\
PGC034671        &BCD   &9.90  &2.22 & 24.49 & 7.67  & $-$    &$-$3.24  &$-$2.81   & 1  & 0 & 0 & 0 & 0 \\
UGC06757         &Irr   &4.61  &1.49 & 25.01 & 7.12  & $-$    &$-$3.78  &$-$3.02   & 1  & 0 & 0 & 0 & 0 \\
Grapes           &Irr   &6.85  &0.54 & 24.24 & 6.54  & $-$    &$-$3.97  &  $-$     & 0  & 0 & 0 & 1 & 0 \\
    \end{tabular}

    \begin{tabular}{llcccccccccccc} \\ \hline
   Name           &  T  & Dist&$A_{26}$ & $SB$ &  $L_K$ & $M_{HI}$& $SFR_{H\alpha}$& $SFR_{FUV}$&   N&  B& F& D& G\\
   \hline
     (1)& (2)& (3)& (4)& (5) & (6)& (7)& (8)& (9)& (10)& (11)& (12)& (13)& (14)\\
   \hline

PGC3401153       &BCD   &9.12  &1.11 & 24.20 & 7.19  & $-$    &$-$3.71  &$-$2.92   & 1  & 0 & 0 & 1 & 0 \\
UGC07320         &Irr   &9.20  &3.15 & 24.33 & 8.04  &7.54  &$-$2.95  &$-$2.61   & 5  & 0 & 0 & 0 & 0 \\
KK135            &Irr   &6.05  &1.18 & 25.84 & 6.58  & $-$    &$-$3.28  &$-$3.10   & 1  & 1 & 0 & 0 & 0 \\
PGC5059199       &Irr   &2.70  &0.09 & 23.59 & 5.25  &6.06  &$-$4.72  &  $-$     & 1  & 0 & 0 & 0 & 0 \\
AGC724906        &Irr   &7.55  &0.85 & 23.86 & 7.09  &7.02  &$-$4.42  &$-$3.18   & 0  & 0 & 0 & 0 & 0 \\
DDO133           &Im    &4.88  &5.03 & 24.84 & 8.24  &8.25  &$-$1.99  &$-$1.73   &18  & 3 & 0 & 0 & 0 \\
PGC041749        &Im    &8.24  &2.14 & 24.83 & 7.51  &7.93  &$-$2.64  &$-$2.11   & 4  & 0 & 0 & 0 & 0 \\
UGC7751          &Im    &7.90  &2.73 & 25.39 & 7.49  &7.58  &$-$3.09  &$-$2.43   & 1  & 0 & 0 & 1 & 0 \\
AGC749241        &Irr   &5.62  &0.56 & 24.53 & 6.46  &6.75  &$-$4.79  &$-$3.54   & 0  & 0 & 0 & 0 & 0 \\
KDG178           &Irr   &7.30  &2.36 & 25.90 & 7.16  &7.69  &$-$3.51  &$-$2.64   & 1  & 0 & 0 & 1 & 0 \\
NGC4656UV        &Irr   &7.98  &13.7 & 25.43 & 8.88  &7.61  &$-$2.31  &$-$0.77   & 9  & 1 & 0 & 0 & 0 \\
UGCA298          &BCD   &11.0  &2.86 & 23.57 & 8.43  &7.29  &$-$1.65  &$-$2.23   & 1  & 0 & 0 & 0 & 1 \\
KKH82            &Tr    &7.48  &2.51 & 25.28 & 7.46  & $-$    &$-$4.33  &$-$3.62   & 0  & 0 & 0 & 0 & 0 \\
PGC2229803       &BCD   &6.58  &1.27 & 23.92 & 7.42  & $-$    &$-$3.27  &$-$2.97   & 1  & 0 & 0 & 0 & 0 \\
LV J1328+4937    &Irr   &8.40  &1.08 & 24.12 & 7.20  & $-$    &$-$2.95  &$-$2.54   & 0  & 0 & 0 & 1 & 0 \\
AGC238890        &BCD   &6.80  &1.64 & 24.76 & 7.30  &6.58  &$-$3.94  &  $-$     & 0  & 0 & 0 & 1 & 0 \\
LV J1342+4840    &Im    &8.40  &1.89 & 24.34 & 7.60  & $-$    &$-$3.37  &  $-$     & 1  & 0 & 0 & 0 & 0 \\
dw1343+58        &BCD   &6.95  &1.91 & 24.22 & 7.65  & $-$    &$-$3.17  &$-$2.56   & 0  & 0 & 0 & 1 & 0 \\
PGC2448110       &Irr   &5.42  &0.36 & 22.74 & 6.80  & $-$    &$-$2.14  &$-$3.04   & 1  & 0 & 0 & 0 & 1 \\
UGC09540         &Irr   &9.30  &2.08 & 24.72 & 7.52  &7.99  &$-$2.06  &$-$2.10   & 6  & 0 & 0 & 0 & 1 \\
NGC6503$-$d1       &Tr    &6.25  &0.96 & 25.73 & 6.44  & $-$    &$-$4.73  &$-$4.09   & 1  & 0 & 0 & 0 & 0 \\
UGCA009          &Irr   &4.92  &3.12 & 25.15 & 7.71  &7.53  &$-$2.90  &$-$2.46   & 2  & 0 & 0 & 0 & 0 \\
UGCA015          &Irr   &3.44  &2.24 & 25.62 & 7.08  &7.06  &$-$4.14  &$-$2.86   & 0  & 0 & 0 & 0 & 0 \\
UGC01104         &BCD   &7.55  &4.06 & 24.16 & 8.35  &8.02  &$-$2.01  &$-$1.70   & 5  & 0 & 0 & 1 & 0 \\
UGC5288          &BCD   &11.5  &4.91 & 23.89 & 8.85  &8.74  &$-$1.32  &$-$1.34   & 5  & 0 & 0 & 0 & 1 \\
UGC08760         &Irr   &3.31  &2.24 & 24.84 & 7.54  &7.33  &$-$3.12  &$-$2.52   & 2  & 0 & 0 & 1 & 0 \\
UGC08833         &Irr   &3.25  &1.12 & 24.24 & 7.41  &7.05  &$-$3.34  &$-$2.85   & 2  & 0 & 0 & 0 & 0 \\
UGC09128         &Irr   &2.30  &1.15 & 24.08 & 7.06  &7.10  &$-$3.97  &$-$2.98   & 3  & 0 & 0 & 0 & 0 \\
DDO210           &Irr   &0.98  &0.76 & 24.55 & 6.78  &6.46  &$-$5.34  &$-$3.86   & 1  & 0 & 0 & 0 & 0 \\
DDO216=Pegasus   &Irr   &0.97  &1.60 & 25.34 & 7.35  &6.74  &$-$4.37  &$-$3.54   & 0  & 0 & 0 & 0 & 0 \\
ESO347$-$017       &Im    &7.60  &4.10 & 24.59 & 8.17  &8.15  &$-$1.77  &$-$1.70   & 6  & 0 & 0 & 1 & 0 \\
ESO348$-$009       &Irr   &11.5  &7.45 & 25.12 & 8.47  &8.64  &$-$2.06  &$-$1.60   & 3  & 0 & 0 & 0 & 0 \\
ESO473$-$024       &Irr   &9.90  &4.44 & 25.87 & 7.72  &8.18  &$-$2.10  &$-$1.94   & 4  & 0 & 0 & 0 & 0 \\
AM0106$-$382       &Irr   &8.20  &1.71 & 24.11 & 7.60  &7.58  &$-$2.33  &$-$2.30   & 4  & 0 & 0 & 0 & 0 \\
UGC07866         &Irr   &4.57  &6.92 & 25.91 & 8.19  &7.90  &$-$1.89  &$-$1.75   & 4  & 0 & 1 & 0 & 0 \\
NGC2366          &Im    &3.28  &6.87 & 24.39 & 8.70  &8.63  &$-$0.95  &$-$1.12   &20  & 0 & 0 & 1 & 1 \\
UGC7559          &Irr   &4.97  &4.14 & 24.98 & 8.09  &8.03  &$-$2.06  &$-$1.85   & 9  & 1 & 0 & 1 & 0 \\
UGC08024         &Irr   &4.04  &2.50 & 24.31 & 7.59  &8.28  &$-$2.52  &$-$1.91   & 8  & 0 & 0 & 1 & 0 \\
UGC05340         &Irr   &12.7  &7.23 & 24.77 & 8.40  &8.95  &$-$1.51  &$-$1.04   &10  & 0 & 0 & 1 & 0 \\
UGC08320         &Irr   &4.25  &4.70 & 24.43 & 8.13  &8.33  &$-$2.13  &$-$1.77   &20  & 1 & 0 & 1 & 0 \\
WLM              &Im    &0.98  &3.25 & 24.78 & 7.70  &7.84  &$-$2.68  &$-$2.23   &12  & 1 & 0 & 0 & 0 \\
LGS$-$3            &Tr    &0.65  &0.46 & 25.39 & 5.96  &5.02  &$-$6.85  &$-$5.23   & 0  & 0 & 0 & 0 & 0 \\
NGC1705          &BCD   &5.73  &3.71 & 23.03 & 8.62  &8.06  &$-$0.98  &$-$0.98   & 1  & 0 & 1 & 1 & 1 \\
ESO409$-$015       &Im    &8.71  &3.00 & 24.08 & 8.10  &8.10  &$-$1.47  &$-$1.72   & 6  & 0 & 0 & 1 & 1 \\
DDO043           &Irr   &10.5  &4.47 & 24.34 & 8.13  &8.38  &$-$1.90  &$-$1.71   & 3  & 0 & 0 & 0 & 0 \\
DDO046           &Irr   &10.4  &5.61 & 24.13 & 8.47  &8.58  &$-$2.00  &$-$1.50   & 2  & 0 & 0 & 0 & 0 \\
DDO64            &Irr   &10.9  &6.83 & 24.65 & 8.41  &8.61  &$-$1.30  &$-$1.12   & 4  & 0 & 0 & 0 & 0 \\
KK78             &Irr   &10.9  &1.29 & 24.27 & 7.29  &7.73  &$-$2.76  &$-$2.51   & 0  & 0 & 0 & 1 & 0 \\
DDO70=SexB       &Irr   &1.43  &2.55 & 24.27 & 7.83  &7.71  &$-$2.88  &$-$2.37   & 4  & 0 & 0 & 0 & 0 \\
DDO75=SexA       &Irr   &1.45  &2.56 & 24.22 & 7.57  &7.90  &$-$2.33  &$-$1.92   & 4  & 0 & 1 & 0 & 0 \\
DDO155           &Irr   &2.19  &1.08 & 24.45 & 6.84  &6.92  &$-$2.67  &$-$2.62   & 1  & 1 & 0 & 1 & 1 \\
DDO167           &Irr   &4.25  &1.33 & 24.21 & 7.19  &7.20  &$-$2.87  &$-$2.49   & 2  & 0 & 0 & 0 & 0 \\
DDO169           &Irr   &4.41  &3.65 & 25.32 & 7.73  &7.75  &$-$2.70  &$-$2.22   & 2  & 0 & 0 & 0 & 0 \\
ESO410$-$005       &Tr    &1.93  &0.91 & 24.53 & 6.89  &5.91  &$-$6.26  &$-$3.96   & 0  & 0 & 0 & 0 & 0 \\
ESO540$-$030       &Tr    &3.56  &1.30 & 25.39 & 6.85  &5.99  &$-$5.72  &$-$3.83   & 0  & 0 & 0 & 0 & 0 \\
ESO294$-$010       &Tr    &2.03  &0.88 & 25.01 & 6.30  &5.52  &$-$4.28  &$-$3.81   & 0  & 0 & 0 & 0 & 0 \\
ESO540$-$032       &Irr   &3.63  &1.62 & 25.92 & 6.83  &6.03  &$-$4.98  &$-$3.91   & 1  & 0 & 0 & 0 & 0 \\
ESO379$-$007       &Irr   &5.45  &1.56 & 24.88 & 7.35  &7.54  &$-$3.62  &  $-$     & 1  & 0 & 0 & 0 & 0 \\
ESO321$-$014       &Irr   &3.33  &1.44 & 24.30 & 7.29  &7.25  &$-$3.06  &$-$2.74   & 3  & 0 & 0 & 1 & 0 \\
ESO381$-$020       &Irr   &5.47  &4.59 & 25.09 & 7.95  &8.37  &$-$1.75  &$-$1.77   & 8  & 0 & 0 & 1 & 1 \\
UGCA319          &Irr   &5.75  &2.24 & 23.94 & 8.02  &7.76  &$-$3.30  &$-$2.29   & 1  & 0 & 0 & 0 & 0 \\
ESO269$-$037       &Irr   &3.15  &1.23 & 24.99 & 6.96  &5.51  &$-$4.40  &$-$3.71   & 1  & 0 & 0 & 0 & 0 \\
KK200            &Irr   &4.76  &1.59 & 24.43 & 7.31  &6.86  &$-$4.26  &$-$3.09   & 1  & 0 & 0 & 0 & 0 \\
ESO384$-$016       &BCD   &4.49  &1.82 & 24.14 & 7.83  &6.69  &$-$4.65  &$-$3.36   & 0  & 0 & 0 & 1 & 0 \\
Cen6             &Irr   &5.94  &1.70 & 24.48 & 7.49  &7.65  &$-$3.26  &$-$2.56   & 1  & 0 & 0 & 1 & 0 \\
UGCA365          &Irr   &5.42  &2.54 & 24.92 & 7.73  &7.29  &$-$4.42  &$-$2.77   & 0  & 0 & 0 & 0 & 0 \\
ESO444$-$084       &Irr   &4.61  &1.91 & 24.16 & 7.61  &7.90  &$-$2.53  &$-$2.28   & 3  & 0 & 0 & 0 & 0 \\
NGC5237          &BCD   &3.33  &2.10 & 23.10 & 8.43  &7.48  &$-$2.26  &$-$2.38   & 0  & 1 & 0 & 1 & 1 \\
IC4316           &Irr   &4.35  &2.67 & 24.57 & 8.21  &7.04  &$-$2.63  &$-$2.38   & 4  & 0 & 0 & 1 & 0 \\
ESO325$-$011       &Irr   &3.40  &3.30 & 24.88 & 7.86  &7.87  &$-$1.88  &  $-$     & 3  & 1 & 0 & 1 & 0 \\
NGC5408          &Im    &5.32  &4.50 & 22.85 & 8.57  &8.48  &$-$0.83  &  $-$     &17  & 0 & 0 & 0 & 1 \\
ESO223$-$009       &Irr   &6.31  &6.13 & 23.93 & 9.17  &8.79  &$-$1.00  &  $-$     & 8  & 0 & 0 & 0 & 0 \\
CGCG014$-$054      &Im    &9.60  &3.53 & 24.66 & 8.01  &7.70  &$-$3.61  &$-$2.32   & 2  & 0 & 0 & 0 & 0 \\
ESO572$-$034       &BCD   &10.5  &4.07 & 23.35 & 8.76  &8.30  &$-$1.04  &  $-$     & 4  & 0 & 0 & 0 & 1 \\
HIPASS J1258$-$04  &Irr   &6.50  &1.59 & 23.40 & 7.82  &7.67  &$-$2.45  &$-$2.24   & 6  & 0 & 0 & 1 & 0 \\
LV J0935$-$1348    &BCD   &9.06  &1.13 & 23.20 & 7.60  &7.29  &$-$2.58  &$-$2.58   & 3  & 0 & 0 & 0 & 1 \\
 \end{tabular}

 \begin{tabular}{llcccccccccccc} \\ \hline
Name           &  T  & Dist&$A_{26}$ & $SB$ &  $L_K$ & $M_{HI}$& $SFR_{H\alpha}$& $SFR_{FUV}$&   N&  B& F& D& G\\
\hline
  (1)& (2)& (3)& (4)& (5) & (6)& (7)& (8)& (9)& (10)& (11)& (12)& (13)& (14)\\
\hline

LV J0944$-$2254    &BCD   &8.20  &0.80 & 22.76 & 7.48  &7.20  &$-$2.59  &  $-$     & 1  & 0 & 0 & 0 & 0 \\
MCG $-$03$-$34$-$002   &BCD   &7.90  &2.99 & 23.64 & 8.10  &7.72  &$-$2.35  &$-$1.86   & 1  & 0 & 0 & 1 & 0 \\
NGC4765          &BCD   &5.65  &3.14 & 23.21 & 8.49  &7.96  &$-$1.56  &$-$1.58   & 7  & 0 & 0 & 0 & 1 \\
UGC06014         &Im    &17.0  &5.99 & 23.87 & 8.78  &8.22  &$-$1.95  &$-$1.53   &11  & 0 & 0 & 0 & 0 \\
UGC07512         &Irr   &11.9  &4.94 & 24.53 & 8.35  &8.36  &$-$2.35  &$-$1.97   & 7  & 0 & 0 & 0 & 0 \\
UGC07596         &Irr   &4.60  &2.09 & 24.41 & 7.65  &6.26  &$-$4.06  &$-$3.21   & 2  & 0 & 0 & 0 & 0 \\
UGC07636         &Irr   &5.20  &1.36 & 23.21 & 7.76  &6.17  &$-$3.50  &$-$3.35   & 0  & 0 & 0 & 0 & 0 \\
VCC114           &Irr   &8.20  &1.42 & 23.60 & 7.64  &7.53  &$-$3.13  &$-$2.78   & 2  & 0 & 0 & 0 & 0 \\
VCC565           &Irr   &9.60  &1.61 & 23.19 & 7.91  &7.06  &$-$3.34  &$-$2.85   & 0  & 0 & 0 & 1 & 0 \\
VCC1675          &BCD   &7.50  &3.09 & 24.08 & 8.22  &6.77  &$-$3.14  &$-$2.51   & 4  & 0 & 0 & 0 & 0 \\
VCC1713          &Im    &7.20  &1.21 & 23.19 & 7.66  &7.08  &$-$3.66  &$-$2.60   & 0  & 0 & 0 & 0 & 0 \\
VCC1725          &BCD   &8.50  &2.90 & 23.76 & 8.48  &7.60  &$-$2.10  &$-$1.66   &14  & 0 & 0 & 1 & 0 \\
VCC2033          &BCD   &9.80  &1.59 & 22.57 & 8.32  &7.01  &$-$2.29  &$-$2.11   & 2  & 0 & 0 & 0 & 1 \\
VCC2037          &Irr   &9.64  &3.35 & 24.70 & 7.94  &6.90  &$-$2.54  &$-$2.33   & 3  & 0 & 0 & 0 & 0 \\
NGC4163          &Im    &2.99  &1.93 & 23.90 & 7.92  &7.17  &$-$2.96  &$-$2.40   & 7  & 1 & 0 & 1 & 0 \\
NGC4190          &BCD   &2.83  &1.80 & 23.70 & 7.90  &7.47  &$-$2.42  &$-$2.22   & 5  & 0 & 0 & 1 & 0 \\
NGC4068          &Im    &4.39  &3.74 & 24.07 & 8.30  &8.08  &$-$1.82  &$-$1.65   & 4  & 1 & 0 & 1 & 0 \\
\end{tabular}
\end{center}

%% file: Karachentsev.bbl
\begin{thebibliography}{}
 


\bibitem{}Bell E.F., McIntosh D.H., Katz N., Weinberg M.D., 2003, ApJS, 149, 289
\bibitem{}Bouchard A., Da Costa G., Jerjen H., 2009, AJ, 137, 3038
\bibitem{}Cote S., Draginda A., Skillman E.D., Miller B.W., 2009, AJ, 138, 1037
\bibitem{}Courtes G., Petit H., Sivan J.P., Dodonov S., Petit M., 1987, A \& A, 174, 28
\bibitem{}Dale D.A., Cohen S.A., Johnson L.C., et al. 2009, ApJ, 703, 517
\bibitem{}Ercan E.N., Aktekin E., Cesur N., Tumer A., 2018, MNRAS, 481, 2804
\bibitem{}Evans C., Castro N., Gonzalez O., et al, 2019, A \& A, 622A, 129
\bibitem{}Hodge P., Miller B.W., 1995, ApJ, 451, 176
\bibitem{}Hunter D.A., Hawley W.N., Gallagher J.S., 1993, AJ, 106, 1797
\bibitem{}James P.A., Shane N.S., Beckman J.E., et al. 2004, A \& A, 414, 23
\bibitem{}Jarrett T.N., Chester T., Cutri R., et al. 2003, AJ, 125, 525
\bibitem{}Jarrett T.N., Chester T., Cutri R., et al. 2000, AJ, 119, 2498
\bibitem{}Kaisin S.S., Karachentsev I.D., 2019, AstBu, 74, 1
\bibitem{}Karachentsev I.D., Kaisina E.I., 2019, AstBu, 74, 111  
\bibitem{}Karachentsev I.D., Kaisin S.S., Kaisina E.I., 2015, Astrophysics, 58, 487
\bibitem{}Karachentsev I.D., Kaisina E.I., 2013, AJ, 146, 46
\bibitem{}Karachentsev I.D., Makarov D.I., Kaisina E.I., 2013, AJ, 145, 101 (UNGC)
\bibitem{}Karachentsev I., Kaisina E., Kaisin S., Makarova L., 2011, MNRAS, 415L,31
\bibitem{} Kennicutt, R.C., Lee J.C., Funes J.G., et al. 2008, ApJS, 178, 247
\bibitem{}Kennicutt, R.C. 1998, ARA\&A, 36, 189
\bibitem{}Lee, J. C., Gil de Paz, A., Kennicutt, R. C., et al. 2011, ApJS, 192, 6
\bibitem{}Lee J.C., Kennicutt R.C., Funes J.G., Sakai S., Akiyama S., 2009, ApJ, 692, 1305
\bibitem{}Long K.S., Winkler P.F., Blair W.P., 2019, ApJ, 875, 85
\bibitem{}Martin D.C., Fanson J., Schiminovich D., et al, 2005, ApJ, 619, L1
\bibitem{}McGaugh S.S., Schombert J.M., 2014, AJ, 148, 77
\bibitem{}McQuinn K.B.W., van Zee L., Skillman E.D., 2019, arXiv:1910.04167
\bibitem{}McQuinn K.B.W., Skillman E.D., Cannon J.M., et al., 2009, ApJ, 695, 561
\bibitem{}Moumen I., Robert C., Devost D., et al. 2019, arXiv:1909.00766
\bibitem{}Rosenberg J.L., Ashby M.L.N., Salzer J.J., et al. 2006, ApJ, 636, 742
\bibitem{}Salzer J.J., Gronwall C., Lipovetsky V.A., et al. 2000, AJ, 120, 80
\bibitem{}Schlegel D.N., Finkbeiner D.P., Davis M., 1998, ApJ, 500, 525
\bibitem{}Schulte-Ladbeck R.E., Hopp U., 1998, AJ, 116, 2886
\bibitem{}Skillman E.D., 2005, New AR, 49, 453
\bibitem{}Skillman E.D., Cote S., Miller B.W., 2003, AJ, 125, 593
\bibitem{}Stinson G.S., Dalcanton J.J., Quinn T., et al., 2007, ApJ, 667, 170
\bibitem{}van Zee L., 2000, AJ, 119, 2757
\bibitem{}Verheijen M.A.W., 2001, ApJ, 563, 694
\bibitem{}Vucetic M.M., Onic D., Petrov N., et al. 2019, Serb. Astron. J., 198, 1
\bibitem{}Williams B.F., Hillis T.J., Blair W.P. et al, 2019, ApJ, 881, 54

\end{thebibliography}
